\documentclass[lettersize,journal]{IEEEtran}
\usepackage{array}
\usepackage[caption=false,font=normalsize,labelfont=sf,textfont=sf]{subfig}
\usepackage{url}
\usepackage{amsmath,amssymb,amsfonts}
\usepackage{algorithmic}
\usepackage{graphicx}
\usepackage{gensymb}
\usepackage{textcomp}
\usepackage{xcolor}
\usepackage{soul}
\usepackage{verbatim}
\usepackage{mathtools} 
\usepackage{amsmath} 
\usepackage{booktabs}
\usepackage{tabularx}
\usepackage{multirow}
\usepackage{array}
\usepackage{algorithm}
\usepackage[table]{xcolor}

\usepackage{siunitx}
\usepackage[numbers,sort&compress]{natbib}
\usepackage[normalem]{ulem}
\usepackage{subfig}
\usepackage{stfloats}
\def\BibTeX{{\rm B\kern-.05em{\sc i\kern-.025em b}\kern-.08em
    T\kern-.1667em\lower.7ex\hbox{E}\kern-.125emX}}
\usepackage{balance}
\newcommand{\inv}{^{\raisebox{.2ex}{$\scriptscriptstyle-1$}}}

\begin{document}


\title{Modeling and Modulation Optimization for OWC Limited by Electronic and Photonic Bandwidth}

\author{Xiaochen~Liu,~\IEEEmembership{Member,~IEEE,}
        Jean-Paul~M.~G.~Linnartz,~\IEEEmembership{Fellow,~IEEE}%
        ~and~Thiago~E.~B.~Cunha,~\IEEEmembership{Member,~IEEE}
}

%
%

\markboth{Journal of \LaTeX\ Class Files,~Vol.~14, No.~8, August~2015}%
{Shell \MakeLowercase{\textit{et al.}}: Bare Demo of IEEEtran.cls for IEEE Journals}
%


\maketitle

\begin{abstract}
In contrast to radio frequency (RF), where the modulation bandwidth is restricted by regulations to avoid interference, the available bandwidth in optical wireless communication (OWC) is primarily constrained by system components. 
To investigate their frequency characteristics,
we review the bandwidth limitations of components in the PHY layer of OWC links. 
Such limitations typically contribute to a decay in the frequency profile of the gain-to-noise ratio (GNR), which can be modeled by a pole-zero transfer function that is generally low-pass.
To boost performance, we optimize the signal power spectral density (PSD) of DC-biased optical orthogonal frequency-division multiplexing (DCO-OFDM) which allows for modulation beyond the 3-dB end-to-end bandwidth. 
We express the Lagrangian-optimized throughput versus the maximum modulation frequency, for an M-zero N-pole low-pass GNR optical link. 
For optimization implementation, we compare a novel Newton-based algorithm with a newly accelerated version of the Hughes-Hartogs (HH) algorithm, to find the (near-) optimal signal spectrum for theoretical, measured and simulated GNR responses.
As demonstrated numerically, employing the proposed multi-stage response model for optimization improves performance in dealing with the successive bandwidth limitations.

\end{abstract}

%
\IEEEpeerreviewmaketitle

\section{Introduction}
\label{sec:intro}


Theoretical limits to the  throughput and capacity expressions are widely used as benchmarks for radio-frequency (RF) communications. The modulation bandwidth is usually limited by adjacent channel interference or regulatory constraints.
Unlike RF, the modulation bandwidth for optical wireless communication (OWC) systems is usually not constrained by regulations. 
Interference between optical signals is limited because light is highly directional, which further eliminates the need for spectral masks.
However, in OWC, the modulation bandwidth is typically limited by the frequency response of the physical components.
Electronic components, such as op-amps, modulator circuits, and parasitic elements, as well as the photonic components, such as light-emitting diodes (LEDs) and photodiodes (PDs), are usually bandwidth-limited which causes the overall gain-to-noise ratio (GNR) to roll off with increasing frequency.
Since OWC relies primarily on a line-of-sight (LoS) path with negligible delay spread, these low-pass characteristics become the dominant factors influencing channel state information (CSI) and ultimately determining the achievable data rate \cite{10892234,channel}.
Therefore, accurate modeling of the frequency-dependent behavior of the system is essential for estimating the real-time CSI, which enables a modulation optimization that adapts the signal to the frequency characteristics of the channel.
In 1968, Robert G. Gallager expressed the capacity of a linear time-invariant (LTI) Additive White Gaussian Noise (AWGN) channel with a non-flat frequency response \cite{gallager1968information}.
These expressions characterize the achievable throughput and offer a systemic framework to optimize the signal spectrum for intensity modulation with direct detection (IM/DD) using optical orthogonal frequency-division multiplexing (OFDM).

Extensive research has been carried out to address frequency limitations in OWC systems. 
LEDs are widely used as light sources and primarily represent the main bottleneck in bandwidth compared to other components \cite{7362097}.
Regarding this, \cite{10315210,8901160,phosphor3,10892234,8214098} provides solutions to optimize the throughput of OFDM with bit and power loading strategies such as waterfilling \cite{gallager1968information}.
With the adoption of high-speed transmitters, e.g., vertical-cavity surface-emitting lasers (VCSELs), receiver-side limitations, such as the area–bandwidth trade-off of detectors and their impact on throughput, have become increasingly prominent and were investigated in \cite{10299714, receiver_noise}.
However, as all the components jointly influence the overall frequency response, estimating system performance may be inaccurate if only part of the limitations is considered, which we quantify later in this paper. 

Bandwidth limitations are not restricted to front-end components.
For example, a comprehensive study on how the bandwidth limitations of the entire OWC channel, encompassing both the hardware and the optical link, affect the performance of on–off keying (OOK) is provided in \cite{9261607}. Even optical phased arrays (OPAs) or reflective intelligent surfaces (RISs) impose bandwidth restrictions \cite{opa,RIS}, though typically in the \si{\giga\hertz} rather than \si{\mega\hertz} range.
In \cite{RIS}, the modulation bandwidth is strictly limited within the flat region of the frequency response. 
However, the throughput can be further improved by optimizing the transmit signal spectrum beyond its flat range \cite{10892234}.
Theoretical analysis of the capacity of band-limited optical intensity channels was first presented in \cite{1291727}, with a discussion on modulation schemes such as pulse amplitude modulation (PAM) and quadrature amplitude modulation (QAM). Throughput expressions of  DC-biased optical OFDM (DCO-OFDM) that mitigate the bandwidth limitations were derived in \cite{7254237,10995233}. 
While these papers contribute with valuable analytical expressions of throughput with bandwidth constraints, the specific low-pass characteristics of the channel resulting from the cascade of low-pass components have not been thoroughly addressed.

The low-pass GNR limitation can be mitigated by OFDM, as it allows for a Lagrangian approach that optimizes the modulation spectrum \cite{receiver_noise,7254237}.
Since this optimization approach addresses the frequency characteristics over a large bandwidth, we focus on DCO-OFDM among other OFDM variants due to its high spectral efficiency \cite{ofdm_comparison}.
Meanwhile, the frequency-dependent optimization emphasizes the requirement for realistic and comprehensive modeling of the channel frequency response that includes all stage components in a system, not only the most limiting factor (e.g., slow power LEDs). 
To the best of our knowledge, this paper is the first to express the optimized throughput of DCO-OFDM for a low-pass channel that captures the bandwidth constraints imposed by the concatenation of various components in an OWC system. In summary, its contributions include:
\begin{itemize}
\item We review bandwidth limitations of components, justifying our model of the OWC link as  cascaded low-pass transfer functions.
\item We optimize the throughput of DCO-OFDM with frequency-limited GNR by Lagrangian waterfilling, resulting in a closed-form  expression for the optimized throughput as a function of the modulation bandwidth.
\item We implement and compare two efficient techniques for the bit-loading algorithm: one leveraging the existence of a maximum modulation frequency, and another accelerating the Hughes-Hartogs (HH) algorithm using channel properties.
\item We discuss a non-monotonic frequency behavior of the channel, as well as their consequences considering the Karush–Kuhn–Tucker (KKT) condition of the optimization.
\item We explore throughput performance using GNR models derived from experiments and circuit simulations, and validate the results by comparison with numerically simulated throughput. By benchmarking the proposed approach against baseline schemes without optimization and against throughput optimization based on inaccurate channel models, we show both the performance gains enabled by optimization and the degradation caused by inaccurate modeling.

\end{itemize}
\begin{table*}[ht]
\centering
\renewcommand{\arraystretch}{1} 
\caption{Comparison Between This Work And Previously Published Work}
\label{tab:related work}
\begin{tabular}{>{\centering\arraybackslash}p{1.5cm} >{\centering\arraybackslash}p{3.5 cm} >{\centering\arraybackslash}p{4cm} >{\centering\arraybackslash}p{1.4cm} >{\centering\arraybackslash}p{3cm} >{\centering\arraybackslash}p{2cm}}
\toprule
\rowcolor{gray!20} 
\textbf{Reference} & \textbf{Frequency Response Model}  & \textbf{Limiting Components} & \textbf{Modulation} & \textbf{Power Loading}  & \textbf{CF Expression$^{1}$} \\
\midrule
 \cite{10315210} & First-order low-pass & LED & OFDM & Power-only & No\\
\midrule
 \cite{8214098} & Experimental data & LED & CAP & Uniform & No\\
\midrule
 \cite{10299714} & First-order low-pass & PDs & OFDM & Uniform & Yes\\
\midrule
 \cite{phosphor3} & High-order low-pass & LED+Phosphor & OFDM & Uniform & No\\
\midrule
 \cite{RIS} & Experimental data & Propagation, Front-ends & PAM & Uniform & Yes\\
\midrule
 \cite{receiver_noise} & First-order low-pass & PDs, op-amp, Noise & PAM,OFDM & Waterfilling & Yes\\
\midrule
 \cite{9261607} & High-order filter & Front-ends & OOK & Uniform & No\\
\midrule
 \cite{1291727} & N.A.$^{2}$ & N.A. & PAM, QAM & Uniform & Yes\\
\midrule
 \cite{7254237} & N.A. & N.A. & OFDM & Waterfilling & Yes\\
\midrule
\textbf{This Work} & High-order filter & Front-ends, Propagation, Noise & OFDM & Waterfilling &Yes\\
\bottomrule
\hline
\end{tabular}
\flushleft
    $^{1}$ Closed-Form Expression, $^{2}$ Non-Applicable
\end{table*}
In Table \ref{tab:related work}, we summarize the related works and compare them with the contributions of this paper. The rest of the paper is organized as follows. Section \ref{sec:h(f)} reviews the bandwidth limitations of OWC system components, which contribute to the spectral GNR modeling in Section \ref{sec:system}. Section \ref{sec:OFDM} derives the throughput expression of DCO-OFDM with the signal spectrum optimized by Lagrangian waterfilling, where the algorithms that implement the optimization to allocate the signal spectrum are shown in Section \ref{sec:algorithm}. Section \ref{sec:condition} discusses the conditions for throughput optimization concerning the GNR frequency characteristics. The throughput enhancement and the algorithm performance are quantified in Section \ref{sec:simu} with channel response and noise reported in experiments or simulations. Section \ref{sec:conclusion} concludes the paper. 

\textit{Notation}:
Signals in the time and frequency domains are denoted by lower-case and upper-case letters with $(t)$ and $(f)$. The imaginary unit is denoted as $j$. Operators $\mathbb{E}[\cdot]$, $|\cdot|$, $[\cdot]^+$ and $\lfloor \cdot \rfloor$ represent respectively, the expectation of a random variable, the absolute value, the non-negative projection, and the floor operation.


\section{Frequency Responses of OWC Link Stages}
\label{sec:h(f)}
\begin{figure*}[t]
  \centering
  \includegraphics[width = 1.65\columnwidth]{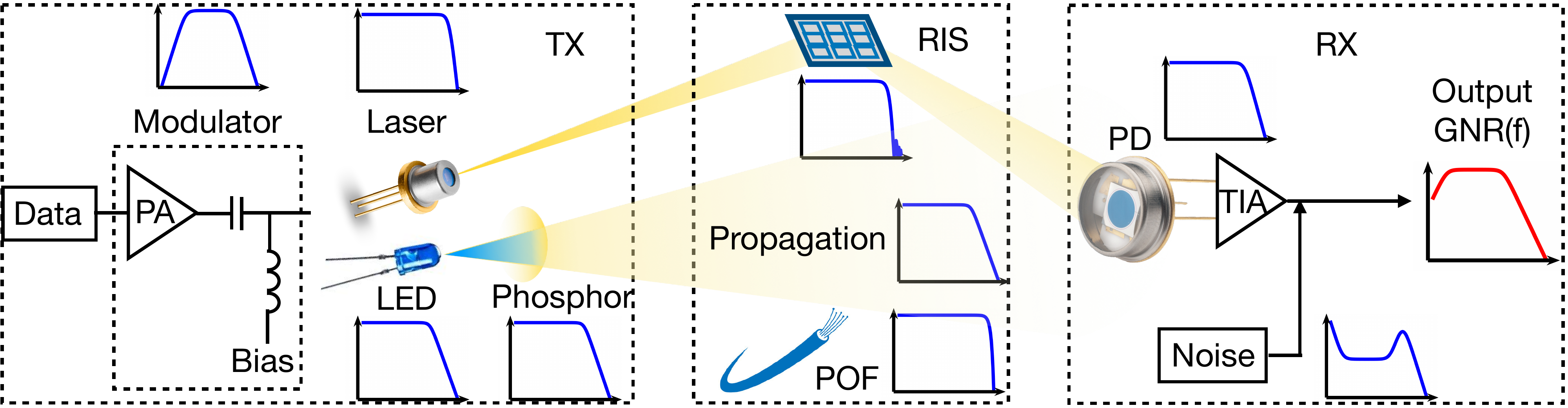}
  \caption{OWC system components and their corresponding frequency characteristics (Response (in blue) and GNR (in red) as a function of frequency).
  }
  \label{fig:OWC link} 
\end{figure*}

IEEE 802.11bb published reference models for the frequency response of optical front-ends, characterizing them by band-pass filters with a finite number of zeros and poles \cite{hinrichs2019ieee}.
Motivated by this and its relevance to system design, this section further contributes to such a framework with an overview of the frequency characteristics of the relevant stages of an OWC system, as summarized in Fig.~\ref{fig:OWC link}.

\subsection{Transmitter}


In an IM/DD OWC system, the optical carrier density is modulated by a real and positive signal $x(t)$. To achieve high throughput, not only a strong signal, thus a high signal-to-noise ratio (SNR) is preferred, but also a high bandwidth is needed. However, the response bandwidth of the light sources and modulators is usually limited.
\subsubsection{LEDs}
The ability of the LED to follow the rapid changes in electrical signals is mainly determined by the carrier recombination lifetime in the active region. As the modulation frequency increases, the recombination rate of electrons and holes in the quantum well becomes insufficient to match the modulation speed, causing the optical output to drop. This is shown analytically in the frequency response with a first-order low-pass characteristic \cite{LED_folp1,LED_folp2,LED_folp3}, such that
\begin{equation}
    H_\mathrm{LED}(f) = \frac{h_{0,\mathrm{LED}}}{1+jf/f_\mathrm{LED}},
    \label{eqn:h(f)LED}
\end{equation}
where $f_\mathrm{LED}$ is the 3-\si{\decibel} bandwidth, and $h_{0,\mathrm{LED}}$ is the DC gain of the LED. In standard large-area LEDs, $f_\mathrm{LED}$ is limited by the relatively long carrier recombination lifetime and parasitic capacitance, resulting in a typical 3-\si{\decibel} bandwidth of only a few \si{\mega\hertz} \cite{normalLEDbw}. For OWC systems targeting short range, small coverage and higher modulation speeds, $\mu$-LED arrays are preferred with 3-\si{\decibel} bandwidths ranging from hundreds of \si{\mega\hertz} to several \si{\giga\hertz} \cite{microled1,microled2}.
Research on LED frequency responses has been extended to include equivalent-circuit analysis incorporating higher-order low-pass modeling \cite{led_ho1}. Building upon the conventional RC model, LEDs have also been represented by more complex RLC circuits with experimental validation, showing that their frequency response can be described by a transfer function featuring multiple poles that represent the low-pass behavior, as well as zeros that slow the roll-off
in the frequency response \cite{led_ho4,led_ho5,led_ho6,diegoLED}.

\subsubsection{Phosphors}
Phosphors are luminescent materials that change the wavelength of emitted light. They are widely used, for example, to produce white illumination by coating blue LEDs \cite{phosphor}. The photoluminescence process of phosphor shows a first-order low-pass characteristic in its frequency response \cite{phosphor1,phosphor2}, which is usually modeled as
\begin{equation}
    H_\mathrm{Ph}(f) = \frac{h_{0,\mathrm{Ph}}}{1+jf/f_\mathrm{Ph}}.
    \label{eqn:h(f)phosphor}
\end{equation}
$h_{0,\mathrm{Ph}}$ and $f_\mathrm{Ph}$ are the gain factor and 3-\si{\decibel} frequency of the phosphor.
Therefore, the frequency response of the emitter is jointly influenced by the diodes and the phosphor. 

\subsubsection{Lasers}
\label{sec:laser}
In addition to LEDs, lasers are promising OWC light sources with much higher bandwidth. The small active volume and low threshold current enable fast carrier dynamics and direct modulation beyond \si{\giga\hertz} \cite{laser1,laser2}. The frequency response of lasers is often modeled by a second-order transfer function as \cite{laser_resp}
\begin{equation}
H_\mathrm{laser}(f) = h_{0,\mathrm{laser}}\frac{f_R^2}{f_R^2-f^2-j\gamma f},
    \label{eqn:h(f)Laser}
\end{equation}
where $h_{0,\mathrm{laser}}$, $f_R$, and $\gamma$ denote the efficiency of electrical-to-optical power conversion, the relaxation oscillation frequency, and the damping factor, respectively. The relaxation oscillation occurs inside the lasers, potentially inducing a resonance peak in its frequency response at $f_R$. By increasing the bias current, $f_R$ can be shifted such that $f_R\leq \gamma/\sqrt{2}$. Under this condition, the frequency response exhibits a smooth second-order low-pass characteristic without a resonant peak. The frequency response can be better understood with an equivalent circuit such as presented in \cite{laser_circuit1,laser_circuit2,laser_circuit3}.

\subsubsection{Electrical Modulator and Driver Circuits}


For IM/DD systems, a modulator driver circuit is required to control the light output of LEDs or laser diodes, enabling data transmission through variations in light intensity. A review of diode driver circuits can be found in \cite{modulator}. Two most commonly used configurations for electrical modulation of the emitter are the bias-T modulator and the series modulator \cite{modulator1}. For example, a bias-T modulator combines a DC bias current, supplied through an inductor, with an AC-coupled data signal through a capacitor. This allows both stable biasing and high-speed modulation of the optical source, where the LC bias-T circuit provides a high-pass frequency response to the AC signal \cite{modulator2}.

The frequency responses of more complex modulator circuits can typically be modeled by a set of zeros and poles \cite{modulator1}. 
Additionally, the wiring between the diode and the driver may introduce an impedance that can limit the bandwidth substantially \cite{diegoLED}.
The zeros in the frequency response usually represent a DC block, which is common in most systems and causes the frequency response to be high-pass at low frequencies. In fact, most systems block DC, for instance in the bias-T and in the receive amplifiers. This may result in baseline wander for PAM signals, unless line codes are used to prevent this. However, for OFDM, the DC block only inhibits the use of a few low subcarrier frequencies, which does not substantially affect throughput. 




\subsection{Propagation Channel}

The LoS optical path is often modeled by the Lambertian radiation of the emitter and the angular sensitivity of the photodetector, along with a frequency non-selective path loss caused by the divergence of the light. However, bandwidth limitations may occur, as discussed in this subsection.



\subsubsection{Diffuse Reflection with Mild Delay Spread}
\label{sec:delay spread}
With modest scattering of light, 
the delay spread is minimal even in indoor environments \cite{channel}. Nevertheless, studies have also examined the diffuse link in OWC channels. The low-pass frequency response of the diffusion in indoor infrared channel is reported in \cite{channel1,channel_folp,channel_folp2}. In \cite{channel_folp}, the multipath dispersion is represented by an exponential-decay impulse response in the time domain, resulting in first-order low-pass frequency behavior. Similarly, \cite{channel_folp2} describes the diffuse link using a low-pass transfer function as
\begin{equation}
    H_\mathrm{diff}(f) = \frac{h_{0,\mathrm{diff}}}{1+jf/f_\mathrm{diff}},
    \label{eqn:h(f)diffuse}
\end{equation}
where $h_{0,\mathrm{diff}}$ and $f_\mathrm{diff}$ are the gain factor $(h_{0,\mathrm{diff}} \leq 1 )$  and the 3-\si{\decibel} bandwidth of the diffuse channel, respectively.

\subsubsection{Plastic optical fiber (POF)}
Offering the benefits of being lightweight, easy to install, and immune to electromagnetic interference (EMI), POFs are commonly used in the optical fronthaul to connect the centralized controller to the optical front-end installed in the ceiling. The POF channel is often modeled as a Gaussian low-pass
filter \cite{pof1,pof2}. 
The Gaussian low-pass filter can be approximated by a cascade of first-order low-pass filters \cite{gaussian_folp1}. This forms a multi-pole system, where the poles are determined by the 3-\si{\decibel} bandwidth of the Gaussian filter, and the number of poles determines the approximation order \cite{gaussian_folp}.


\subsubsection{Beam Squint in OPAs / RISs}
OPAs enable directional beamforming and adaptive spatial control \cite{opa}. By integrating these with optical RISs, the critical issue of link blockage in OWC can be effectively mitigated, as it  significantly enhances the received signal power and link reliability \cite{RIS1}.
In such systems,  a phase mismatch may occur across the modulation bandwidth, which results in a  wavelength-dependent spatial steering of the emitted optical beams.  This effect is known as beam squint \cite{beamsquint1,beamsquint2}. 
\begin{figure}
    \centering
    \includegraphics[width=0.85\linewidth]{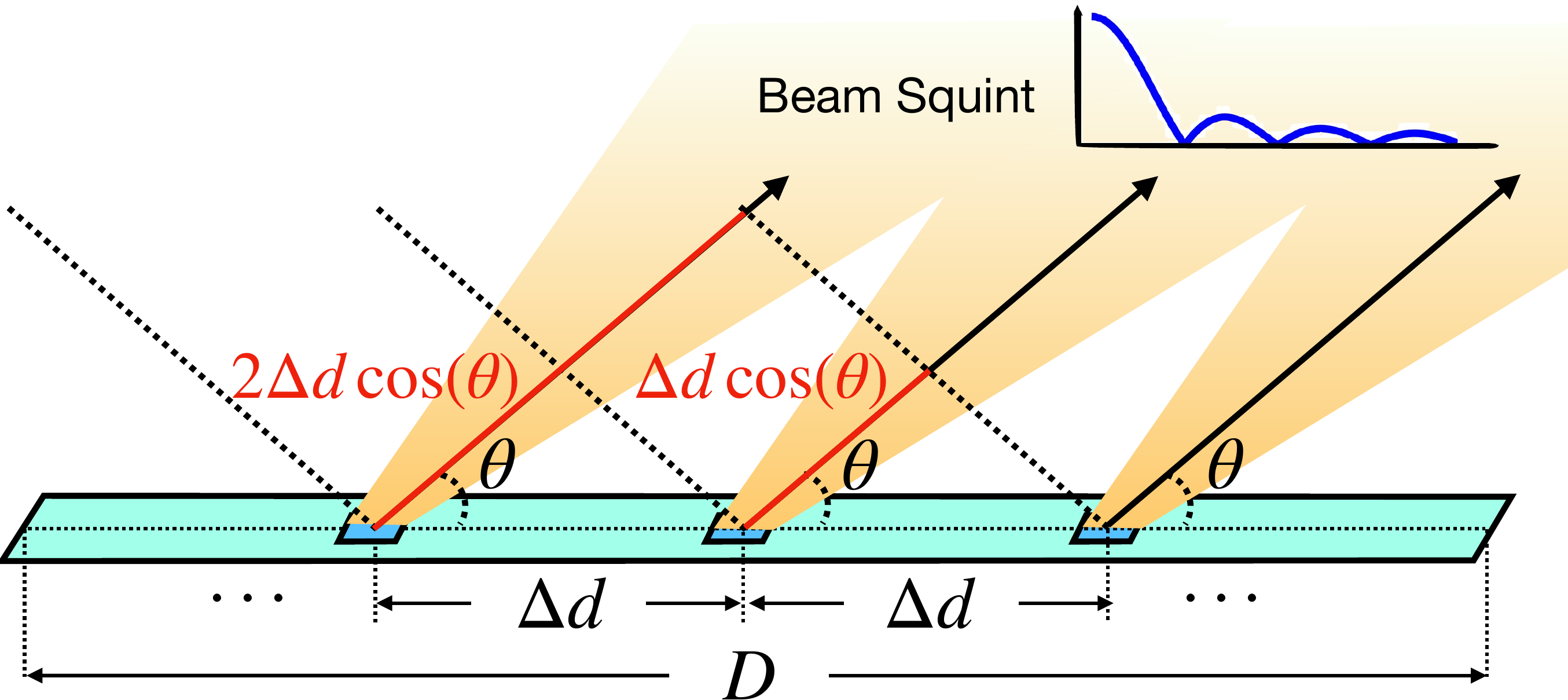}
    \caption{Beam Squint for OPA or RIS.
    Inset: frequency response for modulation}
    \label{fig:squint}
\end{figure}

To achieve phase alignment, the OPAs or RISs compensate for the path length difference between adjacent elements. For an inter-element spacing $\Delta d$, this difference is given by $\Delta d \varphi(\theta)$, where $\varphi(\theta)$ maps the angle of arrival (AoA) or angle of departure (AoD) $\theta$ to the corresponding spatial delay. 
By designing a delay compensation of $\tau_x=x\Delta d \varphi(\theta)/c$ ($c$ is the speed of light) at the $x$-th element, the propagation delays of all $X$ elements are tuned so that all reflected signals arrive at RX with no phase difference.
However, as the intensity of the optical beam is modulated up to a frequency $f$, the propagation phases of the elements deviate from the central frequency according to the modulation frequency. 
As frequency increases, the beam squint of an OPA/RIS link becomes more severe, resulting in a frequency response of a Dirichlet $\mathrm{sinc}$ function, such as 
\begin{equation}
    |H_\mathrm{BS}(f)|= h_{\text{elm}} X\Delta d \varphi(\theta) \left |\mathrm{sinc} \left (\pi f X\Delta d\varphi(\theta)/c \right ) \right |,
\end{equation}
where $h_{\text{elm}}$ is the gain of each element.
The Taylor expansion of the $\mathrm{sinc}$ function $\text{sinc}(x) \approx (1 - x^2/6 + x^4/120-...)$ at $x=0$ provides the polynomial coefficients needed to determine the poles and zeros of a Padé approximation to the original function \cite{baker1996pade}.




\subsection{Receiver}
Achieving high SNR across a wide bandwidth for signals arriving from random directions requires OWC systems to employ detectors with both large aperture and high bandwidth. However, this is challenging due to the electrical properties of PD-based receivers. This subsection  investigates  bandwidth limitations at the receiver.

\subsubsection{PDs and Transimpedance Amplifier (TIA)}
In an OWC receiver using a PD as the detector, the effective PD area affects both the received signal amplitude and the junction capacitance. Although a larger PD area captures more optical power and improves the SNR, it also increases the junction capacitance, thus limiting the receiver bandwidth \cite{PD1}. PD can be represented by an RC-parallel equivalent circuit, resulting in a first-order low-pass frequency response \cite{PD1,PD2} as
\begin{equation}
    H_\mathrm{PD}(f) = \frac{h_{0,\mathrm{PD}}}{1+jf/f_\mathrm{PD}},
    \label{eqn:h(f)PD}
\end{equation}
where $h_{0,\mathrm{PD}}$ is the responsivity of the PD. $f_\mathrm{PD}$ is the PD 3-\si{\decibel} bandwidth which is determined by the junction capacitance. 
As the PD is commonly connected to a high-speed TIA, the overall bandwidth is determined jointly by the PD, the amplifier’s input impedance, and its gain–bandwidth product \cite{PD+tia1,PD+tia2,PD+tia3}. 
In practical implementations, the high-frequency response of the receiver front-end exhibits complex high-order decay. A transfer function with $P_\text{PD-TIA}$ poles is used to analyze the PD–TIA circuit response \cite{receiver_noise}, where the pole frequencies $f_{\mathrm{PD\text{-}TIA},p}$ model the low-pass characteristics determined by the PD capacitance and the op-amp response as
\begin{equation}
    H_\mathrm{PD-TIA}(f) = \frac{h_{0,\mathrm{PD-TIA}}}{\prod_{p=1}^{P_\text{PD-TIA}}(1+jf/f_{\mathrm{PD-TIA},p})},
    \label{eqn:h(f)PD-tia}
\end{equation}
where $h_{0,\mathrm{PD-TIA}}$ includes the PD responsivity and the TIA gain.

\subsubsection{Noise in the TIA}
The dominant noise sources in an OWC system are shot noise from background light and preamplifier noise at the receiver \cite{noisesource1,noisesource2}. The shot noise is generated from the Poisson statistics of photon arrivals from environmental light. With a large light flux, the shot noise generated from the numerous arrivals of photons can be modeled as white Gaussian noise under the central limit theorem. 
However, the amplifier noise at the receiver output is colored. It is amplified at high frequencies due to the combined effects of the amplifier and the PD capacitance, which degrades the spectral GNR especially at high frequencies \cite{PD+tia2}. As frequency further increases, the gain-bandwidth product of the amplifier starts to limit the noise. The limitation, as discussed in the previous subsection, may introduce multiple poles in the transfer function. The amplification and attenuation effects on the noise can be modeled using one zero and multiple poles, as in \cite{receiver_noise}, as
\begin{equation}
    N_\mathrm{amp}(f) = N_{0,\mathrm{amp}}\frac{(1+jf/f_\mathrm{N,PD})}{\prod_{p=1}^{P_N}(1+jf/f_{\mathrm{N,TIA},p})},
    \label{eqn:N(f)}
\end{equation}
where $N_{0,\mathrm{amp}}$ is the noise power spectral density (PSD) at low frequency before amplification. The zero $f_\mathrm{N,PD}$ models the frequency at which the noise begins to be boosted. 
The $P_N$ poles reflect the limitations imposed by the amplifier. Therefore, a subset of these poles is consistent with those in \eqref{eqn:h(f)PD-tia}, as further verified by simulation in Section \ref{sec:simu}.
In addition to noise enhancement at high frequencies, flicker noise dominates at low frequencies \cite{carter2009op}. Its noise spectral density decreases as the frequency increases, thus showing a low-pass frequency characteristic. 
In  OFDM, low-frequency subcarriers can be omitted to mitigate the poor GNR caused by flicker noise.
Since such (near-) DC-block typically only affects a small fraction of the modulation bandwidth, their effect on throughput can be neglected. 


Table \ref{tab:example} summarizes the aforementioned link stages, their frequency-response models, and the corresponding references.

\begin{table*}[ht]
\centering
\renewcommand{\arraystretch}{1.3} 
\caption{Frequency Response of Components in the OWC Link}
\label{tab:example}
\begin{tabular}{>{\centering\arraybackslash}p{3 cm} >{\centering\arraybackslash}p{5cm} >{\centering\arraybackslash}p{6cm} >{\centering\arraybackslash}p{1.5cm}}
\toprule
\rowcolor{gray!20}
\textbf{Components} & \textbf{Frequency Response Characteristics} & \textbf{Transfer Function} & \textbf{References} \\
\midrule
\multirow{3}{*}{LEDs} & First-order Low-pass &
$H_\mathrm{LED}(f) = \frac{h_{0,\mathrm{LED}}}{1 + j f / f_\mathrm{LED}}$ &
\cite{LED_folp1,LED_folp2,LED_folp3} \\[3pt]

 & High-order Low-pass &
$H_\mathrm{LED}(f) = h_{0,\mathrm{LED}}\frac{\prod^{Z_\mathrm{LED}}(1 + j \frac{f}{f_z})}{\prod^{P_\mathrm{LED}}(1 + j \frac{f}{f_p})}$ &
\cite{led_ho1, led_ho4,led_ho5,led_ho6,diegoLED} \\ 
\midrule

Lasers & Second-order Low-pass &
$H_\mathrm{laser}(f) = \frac{f_R^2}{f_R^2-f^2-j\gamma f}$ &
\cite{laser_resp,laser_circuit1,laser_circuit2} \\
\midrule

Phosphor & First-order Low-pass &
$H_\mathrm{Ph}(f) = \frac{h_{0,\mathrm{Ph}}}{1 + j f / f_\mathrm{Ph}}$ &
\cite{phosphor1,phosphor2,phosphor3} \\
\midrule

Modulator Circuit & High-order System &
$H_\mathrm{Mod}(f) = h_{0,\mathrm{Mod}}\frac{\prod^{Z_\text{M}}(1 + j \frac{f}{f_z})}{\prod^{P_\text{M}}(1 + j \frac{f}{f_p})}$ &
\cite{modulator1,modulator2} \\
\midrule

Diffuse  Channel & Approx. First-order Low-Pass & $H_\mathrm{diff}(f) = \frac{h_{0,\mathrm{diff}}}{1+jf/f_\mathrm{diff}}$
&
\cite{channel_folp,channel_folp2}  \\
\midrule

Optical Fiber & Gaussian Low-pass &
$H_\mathrm{POF}(f) = h_{0,\mathrm{POF}} \exp\left \{-\left(\frac{f}{f_\mathrm{POF}} \right)^2\right \}$ &
\cite{pof1,pof2} \\
\midrule

Beam Squint & Sinc Low-Pass & $|H_\mathrm{BS}|= h_{\text{elm}} X\Delta d \varphi(\theta) \left |\mathrm{sinc} \left (\pi f X\Delta d\varphi(\theta)/c \right ) \right |$
&
\cite{beamsquint2}  \\
\midrule

\multirow{1}{*}{PDs} & First-order Low-pass &
$H_\mathrm{PD}(f) = \frac{h_{0,\mathrm{PD}}}{1 + j f / f_\mathrm{PD}}$ &
\cite{PD1,PD2} \\[3pt]

 PD-TIA & Second-order (or higher) Low-pass &
$H_\mathrm{PD-TIA}(f) = \frac{h_{0,\mathrm{PD-TIA}}}{\prod^{P_\text{PD-TIA}}(1+jf/f_\mathrm{PD-TIA,p})}$ &
\cite{receiver_noise,PD+tia1,PD+tia2,PD+tia3} \\ 
\midrule

Receiver Noise & Band-pass &
$N_\mathrm{amp}(f) = N_{0,\mathrm{amp}}\frac{(1+jf/f_\mathrm{N,PD})}{\prod^{P_N}(1+jf/f_{\mathrm{N,TIA},p})}$ &
\cite{receiver_noise,PD+tia2} \\

\bottomrule
\end{tabular}
\end{table*}


\section{System Model}
\label{sec:system}
\begin{figure}[t]
  \centering
  \includegraphics[width = \columnwidth]{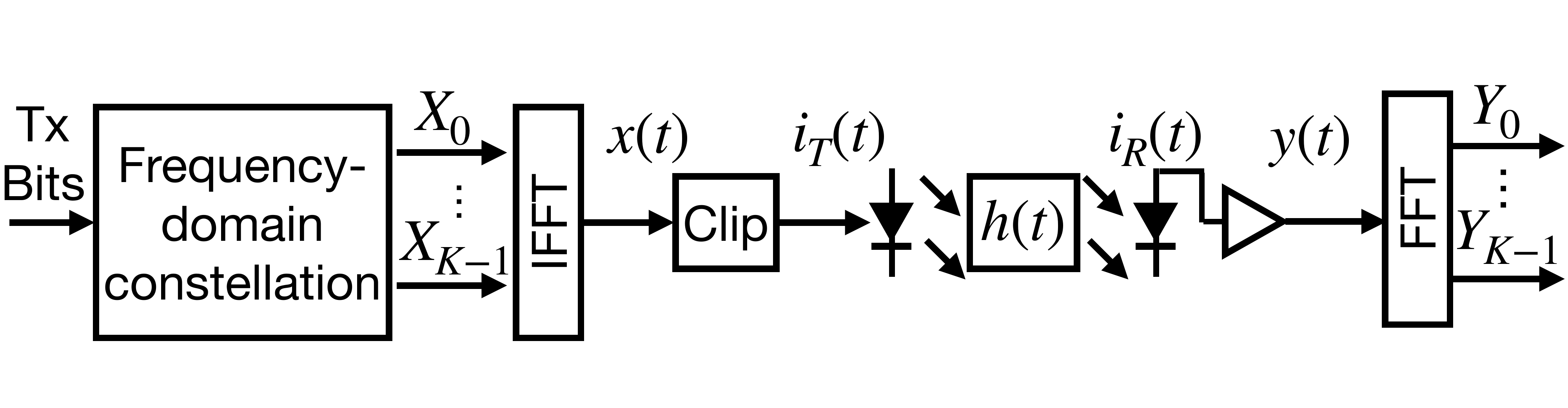}
  \caption{Block diagram of a DCO-OFDM OWC system.}
  \label{fig:OFDM system diagram} 
\end{figure}

In this section, we model the data link where the DCO-OFDM signal undergoes electro-optical conversion, free-space propagation, and opto-electrical conversion.
A block diagram of the system is shown in Fig.~\ref{fig:OFDM system diagram}.
The transmitted bits are mapped to modulated symbols across $K$ subcarriers in the frequency domain, which are then converted into a real-valued time domain signal $x(t)$ via Hermitian symmetry. The signal is fed into the modulator, where it is clipped to a finite range to reduce the peak-to-average power ratio (PAPR),
biased with a DC component, and thereby converted into a non-negative signal $i_T(t)$. 
The current $i_T(t)$ drives the emitter to generate an optical flux, which propagates through free space, potentially after phosphorescence in the phosphor coating.
As mentioned in the previous section, the frequency domain transmit light power density $\Phi_T(f)$ can be expressed as the product of the frequency domain signal $X(f)$ and the frequency response of the modulator, emitter and phosphor, such as
\begin{equation}
    \Phi_T(f) = H_{\mathrm{Mod}}(f)H_{\mathrm{LEDs / lasers}}(f)H_{\mathrm{Ph}}(f) X(f).
\end{equation}
The transmit light flux propagates via free space and is captured by the detector, where the amount of received light power is converted into electrical current by the PD, which is further amplified by the TIA. The output voltage $Y(f)$ of the TIA is denoted as
\begin{equation}
    Y(f) = H_\mathrm{prop}(f)H_\mathrm{PD-TIA}(f) \Phi_T(f).
    \label{eqn:rx}
\end{equation}
In a LoS propagation scenario, the propagation response $H_\mathrm{prop}(f)$ is flat in frequency with amplitude $h_0$ representing the path loss. If POF is used in the optical fronthaul, $H_\mathrm{POF}(f)$ should also be included in \eqref{eqn:rx} \cite{9367435}.

To summarize, the electrical signal $Y(f)$ received by the user device can be expressed in the frequency domain as
\begin{equation}
    Y(f) = X(f) \prod_{ i \in \{  \text{link} \} } H_i(f) + N(f),
\end{equation}
where the product is taken over a subset of $ \text{link} \in \{\text{Mod}, \text{LEDs}/\text{Laser}, \mathrm{Ph}, \text{prop}, \text{PD–TIA}\} $ corresponding to the components included in the particular system configuration.
$X(f)$ is the signal variable in the frequency domain for $f \in [0, f_{\mathrm{max}}] $, where $f_{\mathrm{max}}$ is the maximum modulation frequency. 
We denote the signal PSD as $S(f)$, and the signal variance as $\sigma_x^2$, with the relation of
\begin{equation}
    \sigma_x^2= \int^{f_{\mathrm{max}}}_0 \mathbb{E}[|X(f)|^2] df = \int^{f_{\mathrm{max}}}_0 S(f) df.
\end{equation}
With the signal PSD, the frequency profile of SNR can be expressed as
\begin{equation}
    \mathrm{SNR}(f) =\mathrm{GNR}(f)S(f)= \frac{|H(f)|^2}{S_{N}(f)} S(f),
    \label{eqn:snr}
\end{equation}
where $H(f) = \prod_{ i \in \{  \text{link} \} } H_i(f)$ is the gain response of the overall channel, and $S_N(f)$ is the output noise PSD at the receiver. 
The spectral GNR is defined as the ratio between the power of the total channel frequency response $|H(f)|^2$, and the output noise PSD $S_{N}(f)$.
According to Section \ref{sec:h(f)}, the frequency response of the OWC link can be characterized by the zeros and poles of its transfer function. Therefore, we reformulate $\mathrm{GNR}(f)$ as the product of a DC $\mathrm{GNR}_0$ and a zero-pole transfer function as
\begin{equation}
        \mathrm{GNR}(f)  =  \mathrm{GNR}_{0}
\frac{\prod_{m=1}^{M}\left (1 + \frac{f^2}{f_{z,m}^2}\right )}{\prod_{n=1}^{N}\left (1 + \frac{f^2}{f_{p,n}^2}\right )},
    \label{eqn:gnr_npoles}
\end{equation}
where $M$ and $N$ are the numbers of zeros and poles respectively. The spectral GNR characterizes the frequency behavior of the overall OWC channel, which will be referenced for optimizing the signal spectrum $S(f)$ for throughput, as further introduced in the following sections.


\section{Throughput of DCO-OFDM}
\label{sec:OFDM}
The rate that DCO-OFDM achieves depends on the number of bits that can be allocated per subcarrier, which is proportional to the logarithm of the signal-to-noise ratio. 
This SNR depends on the $\mathrm{GNR}(f)$ but also on the  signal spectrum $S(f)$. The latter can be optimized dynamically, adapting to the link attenuation \cite{gallager1968information}. 
Aggregating the payload of subcarriers over the modulation bandwidth ($0,f_\mathrm{max}$) shows that DCO-OFDM can achieve a bit rate of  
\begin{equation}
    R = \int_0^{f_\mathrm{max}} \left \lfloor\log_2 \left ( 1 + \frac{S(f) \mathrm{GNR}(f)}{\Gamma} \right ) \right \rfloor df .
    \label{eqn:maximize_R}
\end{equation}
$\Gamma$ is a modulation gap that depends on the required bit error rate, the tolerated clipping and the DC bias \cite{gamma}. For ease of calculation, we omit the flooring operation  $\lfloor . \rfloor$, and assume a very fine subcarrier grid; thus we integrate rather than sum over frequencies.
The signal PSD is constrained in power by 
\begin{equation}
    \int_{0}^{f_{\mathrm{max}}} S(f)df \leq \sigma_x^2,
    \label{eqn:power constraint}
\end{equation}
where $\sigma_x^2$ denotes the total transmit signal power.
Additionally, the signal PSD is non-negative by its mathematical definition $(S(f) \geq 0)$.
The throughput can be maximized by shaping the signal power spectrum under a total power constraint according to the GNR \cite{gallager1968information}.
Thus, we formulate the throughput optimization problem as
\begin{align}
\label{eqn:optimization problem}
\max_{S(f)} \quad 
& \int_{0}^{f_{\max}}\log_2 \left ( 1 + \frac{S(f) \mathrm{GNR}(f)}{\Gamma} \right ) df, \\
\text{s.t.} \quad
\label{eqn:kkt_condition}
& \int_{0}^{f_{\max}} S(f) df \le \sigma_x^2, \\
& S(f) \ge 0, \quad \forall f \in [0,f_{\max}].
\label{eqn:CS_condition}
\end{align}
To achieve the optimal signal power allocation for throughput, we employ Lagrangian optimization by building the Lagrangian function with  constraints  \eqref{eqn:kkt_condition} and \eqref{eqn:CS_condition}, such that
\begin{equation}
\begin{split}
    \mathcal{L} &= \int_{0}^{f_{\max}}\log_2 \left ( 1 + \frac{S(f) \mathrm{GNR}(f)}{\Gamma} \right ) df \\
    &-\lambda_1 \left (\int_{0}^{f_{\max}} S(f) df - \sigma_x^2\right ) - \lambda_2(f) S(f).
\end{split}
\label{eqn:lagrangian function}
\end{equation}
The derivative of the Lagrangian function \eqref{eqn:lagrangian function} with respect to 
$S(f)$ is set to zero, such that
\begin{equation}
\frac{d \mathcal{L}}{d S(f)}=\frac{1}{\ln 2}
\frac{\mathrm{GNR}(f)}{\Gamma + \mathrm{GNR}(f)\, S(f)}
- \lambda_1
- \lambda_2(f)
= 0.
\end{equation}
This yields the optimal signal spectrum $S_\mathrm{opt}(f)$ as
\begin{equation}
S_{\mathrm{opt}}(f)
=
\frac{1}{(\lambda_1 + \lambda_2(f)) \ln 2}
-
\frac{\Gamma}{\mathrm{GNR}(f)}.
\label{eqn:optimal_Sf}
\end{equation}
A detailed discussion on the complementary slackness  KKT condition \eqref{eqn:CS_condition} for waterfilling PSD optimization is provided in Section \ref{sec:condition}. For a positive signal PSD over the frequency range $[0,f_\mathrm{max}]$, the multiplier for the complementary slackness condition \eqref{eqn:CS_condition}, $\lambda_2(f)$, is zero. At this time, \eqref{eqn:optimal_Sf} leads to the well-known waterfilling interpretation:
\begin{equation}
    S(f)_{\mathrm{opt}} = \left [ v - \frac{\Gamma }{\mathrm{GNR}(f)} \right ]^+, 
\label{eqn: optimizing power profile}
\end{equation}
where $[\cdot]^+$ ensures that the waterfilling-optimized $S_\mathrm{opt}(f)$ is non-negative.
Here, $v=1/(\lambda_1\ln 2)$ is a constant interpreted as the water level related  to the power budget.
As argued in Section \ref{sec:h(f)}, the derivations in this paper focus on channels in which $\mathrm{GNR}(f)$ is monotonically low-pass decreasing, i.e., the term $\Gamma/\mathrm{GNR}(f)$ increases with frequency. 
In this case, we can
redefine this water level $v$ in terms of the highest frequency where the water depth $v-\Gamma/\mathrm{GNR}(f)$ is positive for $f<f_\mathrm{max}$. 
Monotonicity allows us to write
$v=\Gamma/\mathrm{GNR} (f_{\mathrm{max}})$.
This implies that the water level must always be above the inverse spectral GNR. 
In this case, the optimal signal PSD can be expressed as
\begin{equation}
    S_\mathrm{opt}(f) = \frac{\Gamma }{\mathrm{GNR}(f_\mathrm{max})}   - \frac{\Gamma }{\mathrm{GNR}(f)},
\label{eqn:s(f)}
\end{equation}
for $f<f_\mathrm{max}$ and zero otherwise. That is, $f_\mathrm{max}$ becomes a useful optimization parameter which fully determines $S_\mathrm{opt}(f)$ and $\sigma_x^2$, enabling a one-dimensional optimization.

Interestingly, \eqref{eqn:s(f)} prescribes a gently declining PSD above the 3-\si{\decibel} bandwidth of the link, which is in contrast to  a pre-emphasis that would (counterproductively) boost higher frequencies.
With the signal PSD, the transmit signal power can be calculated as
\begin{equation}
    \sigma_x^2 \hspace{-3pt}=\hspace{-3pt} \int\limits_0^{f_\mathrm{max}} \hspace{-5pt}S_\mathrm{opt}(f) df =   \frac{ f_\mathrm{max} \Gamma }{\mathrm{GNR}(f_\mathrm{max})} -
    \int\limits_0^{f_\mathrm{max}} \frac{\Gamma }{\mathrm{GNR}(f)}  df.
    \label{eqn:sigma}
\end{equation}
For the optimal signal spectrum $S_\mathrm{opt}(f)$, the throughput can be calculated for $S(f) = S_\mathrm{opt}(f)$ via \eqref{eqn:maximize_R}.
We observe that the rate expression is not an explicit function of $\sigma_x^2$, but it appears as a function of $f_\mathrm{max}$. In fact, \eqref{eqn:sigma} 
uniquely relates  $\sigma_x^2$ to  $f_\mathrm{max}$  under the monotonicity
conditions for the GNR profile.

Next, we derive the closed-form expression of the optimum throughput with reference to $f_\mathrm{max}$. 
We consider an OWC channel modeled by a spectral GNR with $N$ poles and $M$ zeros in the transfer function, given by
\begin{equation}
    \begin{split}
        \mathrm{GNR}(f) \hspace{-2pt} = \hspace{-2pt} \mathrm{GNR}_{0}
       \prod_{m=1}^M \hspace{-2pt} \left (1 + \frac{f^2}{f_{z,m}^2} \right )\hspace{-2pt}
       \prod_{n=1}^N \hspace{-2pt} \left (1 + \frac{f^2}{f_{p,n}^2} \right )^{\hspace{-2pt}-1},
    \end{split}
    \label{eqn:gnr_n_pole_m_zero}
\end{equation}
where $\mathrm{GNR}_{0}$ is the GNR of the optical link near DC, 
or more precisely the GNR at low frequencies that are not affected by DC-blocking AC-coupling capacitors or flicker noise. 
$f_{p,n}$ and $f_{z,m}$ are the 3-\si{\decibel} frequency turning points of the poles and zeros, respectively.
According to (\ref{eqn: optimizing power profile}), 
the PSD $S_\mathrm{opt}(f)$ is optimized for a maximum modulation frequency $f_{\mathrm{max}}$ as
\begin{equation}
\begin{split}
    S_\mathrm{opt}(f) &=  \frac{\Gamma }{\mathrm{GNR}(f_\mathrm{max})}   - \frac{\Gamma }{\mathrm{GNR}(f)}\\
    &\hspace{-30pt}= \frac{\Gamma}{\mathrm{GNR}_{0}}
        \left \{ \prod_{m=1}^M \left (1 + \frac{f_\mathrm{max}^2}{f_{z,m}^2} \right )^{-1}
       \prod_{n=1}^N \left (1 + \frac{f_\mathrm{max}^2}{f_{p,n}^2} \right )
        \right . \\
    & \hspace{-27pt}\left . - \prod_{m=1}^M \left (1 + \frac{f^2}{f_{z,m}^2} \right )^{-1}
       \prod_{n=1}^N \left (1 + \frac{f^2}{f_{p,n}^2} \right )
    \right \}.
\end{split}
    \label{eqn: allocated power}
\end{equation}
for $f < f_{\mathrm{max}}$ and zero above $f_{\mathrm{max}}$.
Insert (\ref{eqn: allocated power}) to (\ref{eqn:maximize_R}), we can calculate the throughput as
\begin{equation}
\begin{split}
    R
    &= \int^{f_{\mathrm{max}}} \log_2 \left \{ 1 + \frac{S_\mathrm{opt}(f) \mathrm{GNR}(f)}{\Gamma} \right \} df \\
    &\hspace{-10pt}= \frac{1}{\ln 2}\int^{f_{\mathrm{max}}}\hspace{-5pt} \ln
    \left \{ \hspace{-2pt} \frac{\prod_{m=1}^M (f_{z,m}^2 + f^2)\prod_{n=1}^N (f_{p,n}^2 + f_{\mathrm{max}}^2)}
    {\prod_{m=1}^M (f_{z,m}^2 + f_{\mathrm{max}}^2)\prod_{n=1}^N (f_{p,n}^2 + f^2)} \hspace{-2pt}
    \right \} df \\
    &\hspace{-10pt}= \frac{1}{\ln 2}  \sum_{m=1}^M \hspace{-2pt} \left \{ -f_{\mathrm{max}}\ln(f_{z,m}^2 \hspace{-2pt} + \hspace{-2pt} f_{\mathrm{max}}^2)  
    \hspace{-2pt} + \hspace{-2pt} \int^{f_{\mathrm{max}}} \hspace{-5pt}\ln(f^2 \hspace{-2pt} + \hspace{-2pt} f_{z,m}^2)df \right \}\\
    &\hspace{-10pt} + \frac{1}{\ln 2}  \sum_{n=1}^N \hspace{-2pt} \left \{ f_{\mathrm{max}}\ln(f_{p,n}^2 \hspace{-2pt} + \hspace{-2pt} f_{\mathrm{max}}^2)  
    \hspace{-2pt} - \hspace{-2pt} \int^{f_{\mathrm{max}}} \hspace{-5pt}\ln(f^2 \hspace{-2pt} + \hspace{-2pt} f_{p,n}^2)df \right \}.
\end{split}
\label{eqn:R_ofdm}
\end{equation}
For an integral with the form similar to $\int \ln(x^2+a^2)dx$, integration by part technique can be used, so that
\begin{equation}
    \int\ln(x^2+a^2)dx = xln(x^2+a^2) - 2x + 2a \arctan \left (\frac{x}{a} \right ) + c.
    \label{eqn:int_ln(x2+a2)}
\end{equation}
Applying (\ref{eqn:int_ln(x2+a2)}) to (\ref{eqn:R_ofdm}), we obtain the expression for throughput of DCO-OFDM with the PSD optimized by waterfilling power allocation strategy, over an OWC link with GNR$(f)$ modeled by a transfer function with M zeros and N poles as
\begin{equation}
\begin{split}
    R 
    &= \int^{f_{\mathrm{max}}} \log_2 \left ( 1 + \frac{S_\mathrm{opt}(f) \mathrm{GNR}(f)}{\Gamma} \right ) df \\
      &  =  \frac{2}{\ln 2}  \bigg\{  (N   -  M)f_{\mathrm{max}} \bigg.\\
      &\hspace{-3pt}\left .+   \sum_{m=1}^{M} \hspace{-2pt}  f_{z,m}\arctan   \left ( \hspace{-3pt}  \frac{f_{\mathrm{max}}}{f_{z,m}}  \hspace{-3pt}\right) \hspace{-3pt} - \hspace{-3pt}  \sum_{n=1}^{N} \hspace{-2pt}  f_{p,n}\arctan    \left (  \hspace{-3pt} \frac{f_{\mathrm{max}}}{f_{p,n}}   \hspace{-3pt}\right )\hspace{-3pt}
    \right \}.
\end{split}
\label{eqn:arctan_mzero_npole}
\end{equation}
The coefficient of the linear $f_\mathrm{max}$ term is determined by the number of zeros and poles. The low-pass behavior of $\mathrm{GNR}(f)$ indicates that the number of poles is higher than the number of zeros, thus $N-M$ is positive. Each zero or pole contributes to an arctangent term that shows the effect of frequency response increase or decay on throughput. 

Eq. \eqref{eqn:arctan_mzero_npole} gives a more accurate estimation of the throughput lower bound than $R=f_\mathrm{max}\log_2(1+\mathrm{SNR})$ which assumes uniform power loading of $S(f)$ over the modulation bandwidth $f_\mathrm{max}$. As waterfilling OFDM is capable of optimizing throughput over low-pass links, and the static frequency behavior of link components is well understood and modeled as previously reviewed in Section \ref{sec:h(f)}, this expression provides a basis for evaluating and enhancing OWC link performance.


\section{Low-Pass Condition and Water Level}
\label{sec:condition}
\label{sec:convex}
In the previous section, the spectral GNR has been analyzed under the assumption of a monotonic decrease. In practice, this is typically the case since the power gain cannot increase indefinitely with frequency. 
This section investigates the special scenarios of non-monotonically decreasing $\mathrm{GNR}(f)$. We address the limitation of the closed-form expression \eqref{eqn:arctan_mzero_npole} by revisiting the throughput optimization in \eqref{eqn:optimal_Sf} and reinforcing the complementary slackness condition in \eqref{eqn:CS_condition}. 

\begin{figure*}[!t]
    \centering
    \captionsetup[subfloat]{labelfont=scriptsize,textfont=scriptsize}
    \subfloat[\label{fig:3p2z_example}]{%
        \includegraphics[width=0.25\linewidth]{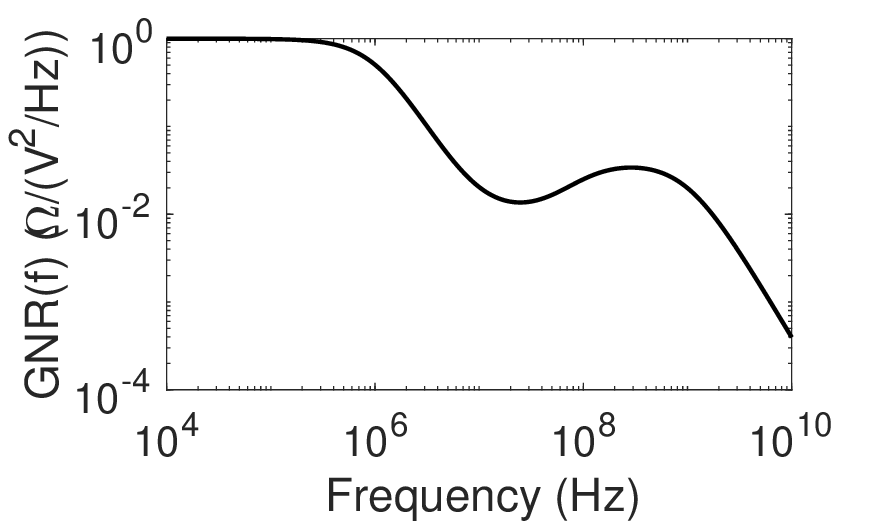}
    }
    \subfloat[\label{fig:fmax_small}]{%
        \includegraphics[width=0.25\linewidth]{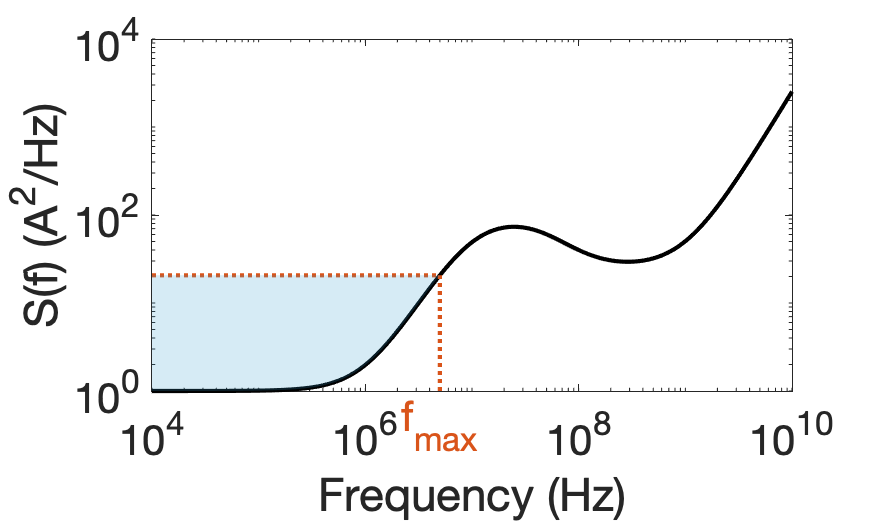}
    } 
    \subfloat[\label{fig:fmax_mid}]{%
        \includegraphics[width=0.25\linewidth]{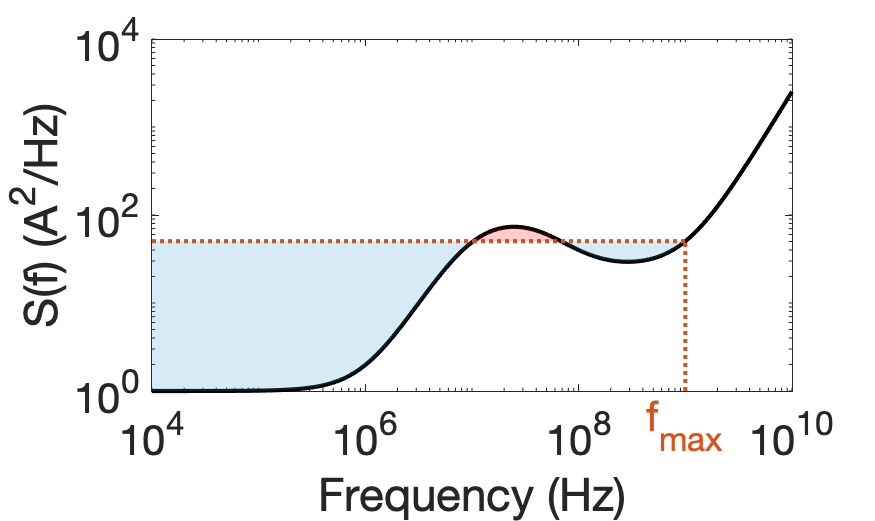}
    } 
    \subfloat[\label{fig:fmax_large}]{%
        \includegraphics[width=0.25\linewidth]{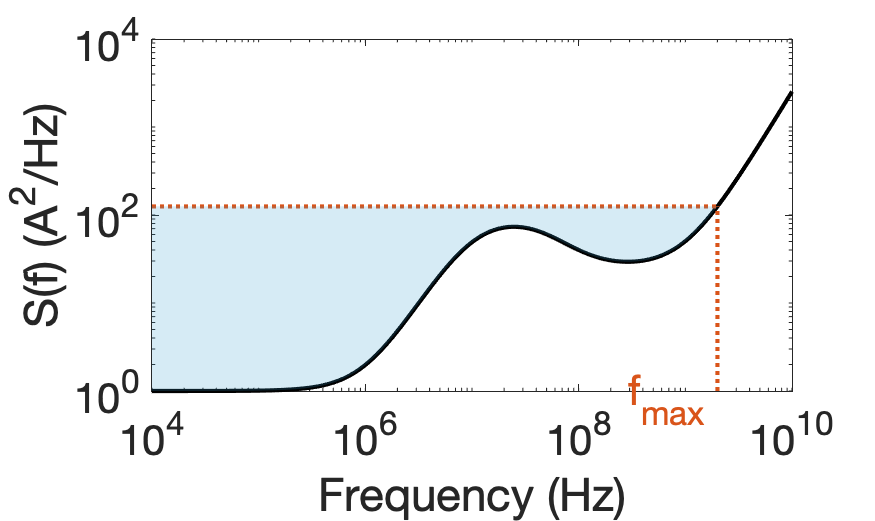}
    }
    \caption{(a) An example of a non-monotonically low-pass $\mathrm{GNR}(f)$ with 3 poles ($f_{p,1} = 1$ \si{\mega\hertz},$f_{p,2} = 100$ \si{\mega\hertz} and $f_{p,3} = 1$ \si{\giga\hertz}) and 2 zeros ($f_{z,1} = 10$ \si{\mega\hertz} and $f_{z,2} = 50$ \si{\mega\hertz}), $\mathrm{GNR}_0 = 1$ \si{\ohm}/(\si{\square\volt}/\si{\hertz}). (b)--(d) show the signal PSD optimization by waterfilling according to $\mathrm{GNR}(f)$, for low, medium, and high power budget. In (c), the complementary slackness condition is highlighted where it constrains the PSD to remain non-negative.}
    \label{fig:fmax}
\end{figure*}

If $\mathrm{GNR}(f)$ does not decrease monotonically, for example in a normalized three-pole, two-zero frequency response as illustrated in Fig.~\ref{fig:3p2z_example}, the resulting non-monotonic behavior can affect the optimization of $S(f)$.
The $\mathrm{GNR}(f)$ curve contains a local minimum region and a peak.
To optimize the signal PSD, $\mathrm{GNR}(f)$ is inverted and used as a 'shape' to allocate the power budget. Within a certain amount of power, $\Gamma/\mathrm{GNR}(f)$ shapes the PSD according to \eqref{eqn:s(f)}, with the water level associated with $f_\mathrm{max}$. As the power budget increases, $f_\mathrm{max}$ will increase accordingly. 

Fig.~\ref{fig:fmax} shows different scenarios for the $S(f)$ profile optimized by waterfilling according to a non-monotonically decreasing $\mathrm{GNR}(f)$. For a small power budget, waterfilling allocates the power according to the low-pass shape of $\mathrm{GNR}(f)$, as illustrated in Fig.~\ref{fig:fmax_small}. For higher frequencies, the zeros in $\mathrm{GNR}(f)$ cause an increase in the response, resulting in a convex region in its inverse curve. 
As the power budget increases and $f_\mathrm{max}$ extends beyond the turning point, the water level $\Gamma/\mathrm{GNR}(f_\mathrm{max})$ is no longer the highest level for $f \leq f_\mathrm{max}$. 
As a result, the optimum solution $S(f)_\mathrm{opt}$ calculated from \eqref{eqn:s(f)} may take negative values, as shown in Fig.~\ref{fig:fmax_mid}. 
In this case, we invoke the complementary slackness condition in \eqref{eqn:CS_condition} to prevent $S(f)_\mathrm{opt}$ from turning negative. 
We define the frequency range of the island region as
\begin{equation}
\mathcal{F}
=
\left\{
f \in [0,f_{\max}] \hspace{-2pt}\;\big|\;\hspace{-2pt}
\Gamma /\mathrm{GNR}(f_\mathrm{max})  \leq \Gamma /\mathrm{GNR}(f)
\right\}.
\end{equation}
Within this frequency range, the multiplier $\lambda_2(f)$ is no longer zero, but becomes positive to activate the constraint \eqref{eqn:CS_condition} as
\begin{equation}
    \lambda_2(f) = \frac{\mathrm{GNR}(f)}{\Gamma\ln 2} - \lambda_1,
\end{equation}
so that 
\begin{equation}
\hspace{-6pt}S(f)_\mathrm{opt} \hspace{-4pt} = \hspace{-4pt}
\begin{cases}
\Gamma \hspace{-2pt}/\mathrm{GNR}(f_\mathrm{max})  \hspace{-3pt} - \hspace{-3pt} \Gamma \hspace{-2pt}/ \mathrm{GNR}(f), 
&\hspace{-5pt} f \hspace{-3pt}\in\hspace{-3pt} [0,f_{\max}] \hspace{-3pt}\setminus\hspace{-3pt} \mathcal{F}, \\[1.5mm]
0, 
& \hspace{-105pt} \text{otherwise} .
\end{cases}
\label{eqn:s(f)_non_mono}
\end{equation}
In this case, $\lambda_2$ compensates for the unconstrained signal spectrum optimization \eqref{eqn:s(f)} from negative values. 
For a sufficiently large power budget and a sufficiently high $f_\mathrm{max}$, such that $\Gamma/\mathrm{GNR}(f_\mathrm{max})$ lies above the local maximum point, the waterfilling solution in \eqref{eqn:s(f)} is valid over the entire frequency range, and the complementary slackness condition in \eqref{eqn:CS_condition} becomes inactive, as shown in Fig.~\ref{fig:fmax_large}.


In a practical optical link, the high-pass uplift in the $\mathrm{GNR}(f)$ is usually near DC as a  result of DC blocking.
As discussed in Section \ref{sec:h(f)}, OFDM can intentionally avoid the use of low-frequency subcarriers and  eliminate the influence of these zeros without a severe penalty for throughput. As frequency increases, it is unlikely that components exhibit a high-boost response, and an unrealistically high power gain.
Yet, the laser frequency response may exhibit a peak when the bias current is insufficient, as reviewed in Section~\ref{sec:laser}. Similarly, at the receiver, the TIA response may also show a resonant peak if the feedback capacitor is not properly selected for compensation \cite{wang1993compensate}. The resonant peak in the frequency response of the transceiver components will result in an 'island' region in Fig.~\ref{fig:fmax_mid}. For the optimization of signal PSD, similar condition applies as discussed previously in \eqref{eqn:s(f)_non_mono}. However, the resonant peak can cause time-domain oscillations, which degrade system stability. Therefore, for components in an OWC link that include feedback systems (e.g., lasers or TIAs), stability compensation is typically required, which flattens the peak in the frequency response. Consequently, the spectral GNR remains monotonically decreasing.


\section{Implementation of Signal Spectrum Optimization}
\label{sec:algorithm}

Eq. \eqref{eqn:arctan_mzero_npole} models the relation between the maximum modulation frequency $f_\mathrm{max}$, which determines the water level, and the estimated throughput. However, the total electrical power needed to achieve such throughput is not derived.
In practice, the system transmits at the maximum allowable power to achieve higher throughput, which makes it critical to derive the relation between achievable throughput and power consumption.
For an optical transmitter, the relationship between electrical power consumption $P_T$ and signal variance $\sigma_x^2$ is generally nonlinear due to the diode nonlinear V–I characteristics \cite{10882056}. 
Here, we express the electrical power consumption as a function of signal variance as
\begin{equation}
    P_T = g(\sigma_x^2).
    \label{eqn:PT_sigma}
\end{equation}
Therefore, the signal variance can be expressed as $g\inv(P_T)$.
In \eqref{eqn:s(f)}, the signal PSD $S(f)$ is optimized according to $\mathrm{GNR}(f)$ given a certain water level. 
For a $n$-pole $m$-zero $\mathrm{GNR}(f)$, it is difficult to express $f_\mathrm{max}$ as a function of $\sigma_x^2$, and furthermore the throughput as a function of $\sigma_x^2$. This complexity impedes the dynamic allocation of the signal spectrum under real-time signal variance constraints.
To efficiently update the new $f_\mathrm{max}$ and the corresponding $S(f)$ as $\sigma_x^2$ changes, we 
\begin{enumerate}
    \item apply Newton's method (NT) to search for the maximum modulation frequency of the waterfilling-optimized signal spectrum for a given signal variance, as shown in Algorithm \ref{alg:newton};
    \item accelerate the HH algorithm ($\text{HH}_\text{A}$) to iteratively load bits to subcarriers up to the power constraint, as shown in Algorithm \ref{alg:HH}. 
\end{enumerate}

We first reformulate the waterfilling-optimized throughput expression and the power constraint over $K$ discrete OFDM subcarriers as
\begin{equation}
\begin{split}
    R &= \Delta_{\mathrm{B}}\sum_{k=1}^{K}\log_{2} \left(1 + \frac{\mathrm{GNR}(f_k)}{\Gamma} \sigma_x^2(f_k)   \right), \\
    & s.t. \hspace{4pt} \sigma_x^2(f_k) \geq 0 \hspace{4pt}\text{ and } \hspace{2pt} \sum_{k=1}^K \sigma_x^2(f_k)=\sigma_x^2,
\end{split}
\label{eqn:R_discrete}
\end{equation}
where $f_\text{chip}$ is the maximum modulation bandwidth supported by the chipset, and $\Delta_{\mathrm{B}} = f_\text{chip}/K$ denotes the subcarrier bandwidth. We calculate the signal power assigned to the $k$-th subcarrier as $\sigma_x^2(f_k) = S(f_k)_\mathrm{opt} \Delta_{\mathrm{B}} $.

\subsubsection{Newton's Method (NT)}
We adapt a Newton-based algorithm NT to efficiently determine the optimal maximum modulation frequency, thus the optimal signal spectrum, for a given power budget. It starts with precomputing the $\sigma_x^2$ for the maximum value of $f_\text{chip}$ supported by the chipset. This is achieved by summing the products of subcarrier bandwidth and their respective signal PSD computed from \eqref{eqn:s(f)}.
The calculated set of 
$(\tilde{f}_{\mathrm{max},i=0} =f_\text{chip},\tilde{\sigma}_{x,i=0}^2)$ is used as the starting point at step $i=0$, for the algorithm to iteratively approach the target $(f_\mathrm{\mathrm{max}},\sigma_x^{2})$,
that is, the optimum signal spectrum with maximum modulation frequency $f_\mathrm{\mathrm{max}}$ constrained by a signal variance $\sigma_x^{2}$.
At each step, the distance from $\tilde{f}_{\mathrm{max},i}$ to $f_{\mathrm{max}}$, is determined by the difference between the signal variance $\tilde{\sigma}_{x,i}^2-\sigma_x^{2}$, divided by the derivative of the signal variance with respect to $\tilde{f}_\mathrm{max,i}$ at this point. The derivative of $\sigma_x^2(\tilde{f}_\mathrm{max,i})$ is calculated as
\begin{equation}
\begin{split}
        \frac{\mathrm{d}\tilde{\sigma}_x^2}{\mathrm{d}\tilde{f}_\mathrm{max,i}} & \hspace{-2pt}= \hspace{-2pt}  
        \left[ \int_0^{\tilde{f}_\mathrm{max,i}} \hspace{-4pt}\left(\frac{\Gamma }{\mathrm{GNR}(\tilde{f}_\mathrm{max,i})}   - \frac{\Gamma }{\mathrm{GNR}(f)} \right )df \right]'\\
    &\hspace{-40pt}=\hspace{-2pt}\frac{2\Gamma \tilde{f}_\mathrm{max,i}^2 \hspace{-2pt} \left( \hspace{-2pt} 
    \sum_{n=1}^N\frac{1}{(f_{p,n}^2 + \tilde{f}_\mathrm{max,i}^2)} \hspace{-2pt}-\hspace{-2pt} \sum_{m=1}^M\frac{1}{(f_{z,m}^2 + \tilde{f}_\mathrm{max,i}^2)}  \hspace{-3pt} \right )}{\mathrm{GNR}(\tilde{f}_\mathrm{max,i})}.
\end{split}
\end{equation}
The iterative search stops when the update frequency distance is smaller than the subcarrier 
spacing
and the total signal variance is strictly limited by $\tilde{\sigma}_{x,i}^2 \leq \sigma_x^{2}$, and thus the power consumption $P_T$ via \eqref{eqn:PT_sigma}. We illustrate the algorithm in Algorithm \ref{alg:newton}.
\begin{algorithm}[ht]
\caption{Newton's method (NT) to reach $f_\mathrm{max}$ given $\sigma_x^2$}
\begin{algorithmic}[1]
\STATE \textbf{Initialize:} $\mathrm{GNR}(f)$, maximum modulation frequency supported by the chipset $f_{\text{chip}}$, subcarrier number $ K$, signal variance constraint $\sigma_x^{2}$.
\STATE Set $\Delta_{\mathrm{B}} = f_{\text{chip}}/K$, $W(\cdot) = \Gamma \mathrm{GNR}^{-1}(\cdot)$, $\tilde{f}_{\mathrm{max},0} \leftarrow f_{\text{chip}}$;
\STATE $\tilde{\sigma}_{x,0}^2 =\Delta_{\mathrm{B}} \sum_{k=1}^{ K} \left (W (\tilde{f}_{\mathrm{max},0})-W(k \Delta_{\mathrm{B}})\right)$;
\WHILE{true}
\STATE $\tilde{f}_{\mathrm{max},i+1} = \tilde{f}_{\mathrm{max},i} + (\tilde{\sigma}_{x,i}^2-\sigma_x^2)/\frac{\mathrm{d}\tilde{\sigma}_x^2(\tilde{f}_{\mathrm{max},i})}{\mathrm{d}\tilde{f}_{\mathrm{max},i}}$;
\STATE Search $k^* \leftarrow \arg\min\limits_{k \in \{1, \dots,  K\}} \left| \tilde{f}_{\mathrm{max},i+1}-k\Delta_{\mathrm{B}} \right|$;
\FOR{$k = 1$ \textbf{to} $ k^*$}
\STATE  $S(f_k)=W(k^*\Delta_{\mathrm{B}})-W(k \Delta_{\mathrm{B}})$;
\STATE $\tilde{\sigma}_{x,i+1}^2 \leftarrow \tilde{\sigma}_{x,i}^2+\Delta_{\mathrm{B}}\max[0,S(f_k)]$;
\ENDFOR
\IF{$k^*\Delta_{\mathrm{B}} - \tilde{f}_{\mathrm{max},i} < \Delta_{\mathrm{B}} \ \textbf{and} \ \tilde{\sigma}_{x,i+1}^2 < \sigma_x^2$}
    \STATE Break;
\ENDIF
\STATE $i \leftarrow i+1$;
\STATE $\tilde{f}_{\mathrm{max},i} = k^* \Delta_{\mathrm{B}}$;
\ENDWHILE
\end{algorithmic}
\label{alg:newton}
\end{algorithm}


\subsubsection{Accelerated HH Algorithm ($\text{HH}_\text{A}$)}
As an alternative to Algorithm \ref{alg:newton}, one may apply the HH Algorithm \cite{HughesHartogs1987},
which iteratively loads bits onto the subcarriers that have the highest potential to boost throughput, given the GNR and the amount of power and bits already allocated in previous iterations. 
We calculate the extra rate $\Delta R$ achieved in the $k$-th subcarrier per extra power $\Delta \sigma_x^2$ from \eqref{eqn:R_discrete}, as
\begin{equation}
\begin{split}
    &\frac{\Delta R(f_k)}{\Delta \sigma_x^2(f_k)} = \Delta_{\mathrm{B}} \hspace{-2pt}  \Bigg[\log_2 \hspace{-2pt} \Bigg( \hspace{-2pt} 1+\dfrac{\mathrm{GNR}(f_k) \left(\sigma_x^2(f_k)+\Delta \sigma_x^2(f_k)\right)}{\Gamma} \hspace{-2pt} \Bigg)\\
    &  \hspace{25pt} -\log_2\Bigg(1+\dfrac{\mathrm{GNR}(f_k) S(f_k)}{\Gamma}\Bigg)\Bigg]/\Delta \sigma_x^2(f_k).
    \label{eqn:derivative_R}
    \end{split}
\end{equation}
The ratio $\Delta R(f_k)/\Delta \sigma_x^2(f_k)$ is a discrete version of
the derivative  $d R/d S(f)$ of \eqref{eqn:maximize_R} that accounts for unit steps of one extra bit per OFDM frame.
 We reshuffle \eqref{eqn:derivative_R}, such that


\begin{equation}
    \Delta  \tilde{\sigma}_x^2 {(f_k)} \hspace{-2pt}= \hspace{-1pt}
    \left ( \hspace{-1pt} \tilde{\sigma}_x^2 {(f_k)} \hspace{-2pt}+ \hspace{-2pt} \Delta_{\mathrm{B}} \frac{\Gamma }{\mathrm{GNR}(f_k)} \hspace{-2pt}\right ) \hspace{-3pt}
    \left(\hspace{-1pt} 2^\frac{\Delta R(f_k)}{\Delta_{\mathrm{B}}} \hspace{-2pt} - \hspace{-2pt} 1 \hspace{-1pt} \right).
    \label{eqn:HH}
\end{equation}
By setting the bit-loading increment $\Delta R(f_k)/\Delta_{\mathrm{B}}$ for each subcarrier to one bit, the second term in \eqref{eqn:HH} becomes unity \cite{10295463}.
The subcarrier $k$ that minimizes \eqref{eqn:HH} 
indexes the modulation frequency $f_k$ that has the highest power efficiency. It requires the least amount of power to boost the throughput, thus becoming the candidate to carry the next bit.

\begin{figure}[t]
    \centering
    \captionsetup[subfloat]{labelfont=scriptsize,textfont=scriptsize}
    \subfloat[Subcarrier 1 receives 1 bit  \label{fig:HH_table_1}]{%
        \includegraphics[width=.85\linewidth]{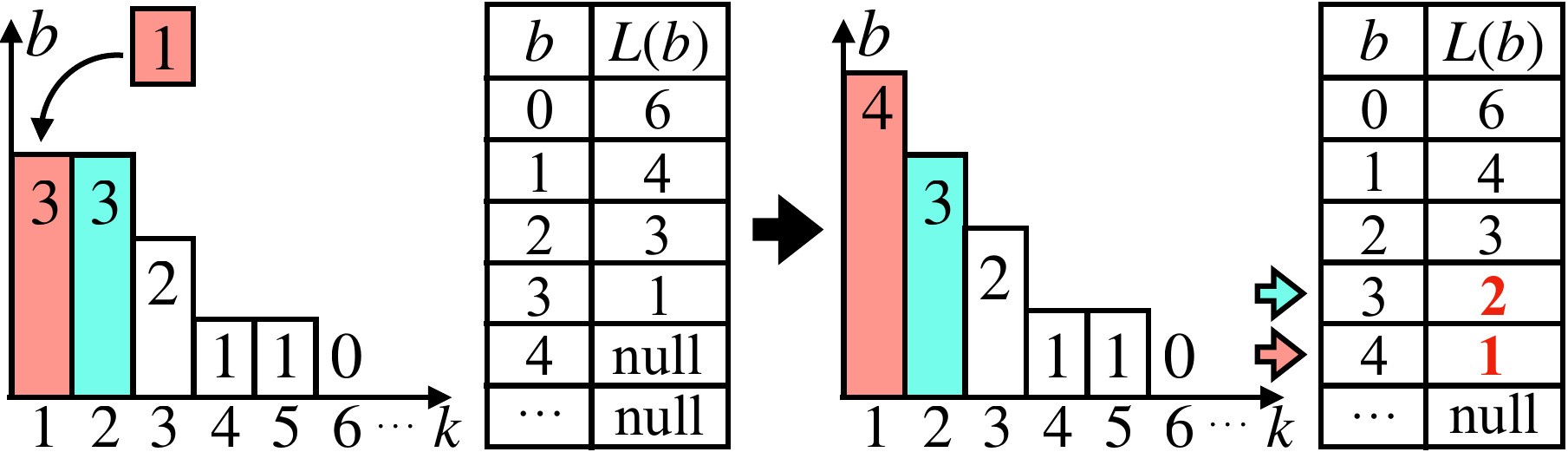}
    }\\
    \subfloat[Subcarrier 3 receives 1 bit \label{fig:HH_table_2}]{%
        \includegraphics[width=.85\linewidth]{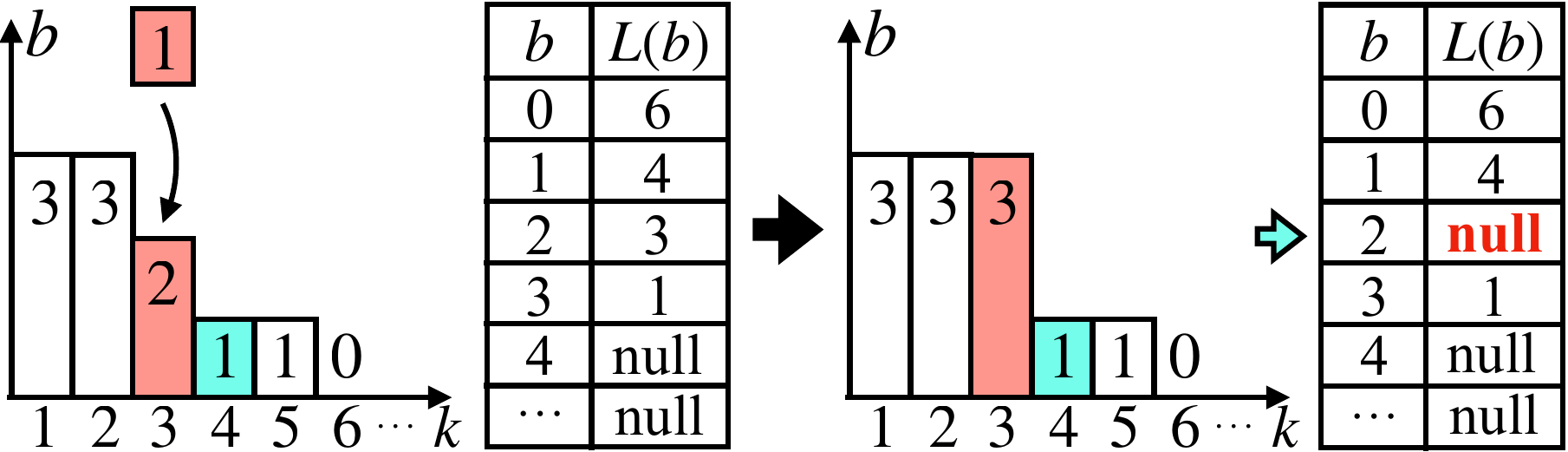}
    }
    \caption{Two bit-loading scenarios and their corresponding lookup table update. The red and blue subcarriers are, the subcarrier that receive one additional bit, and its upper adjacent subcarrier, respectively.}
    \label{fig:HH_table}
\end{figure}

We further exploit the insight that for a monotonically declining GNR$(f)$, 
both $S(f)$ and the bit loading must also be monotonically declining with frequency.
As bit loading makes discrete steps, it shows a staircase profile that decreases monotonically, as illustrated in Fig.~\ref{fig:HH_table}.
The set of subcarriers carrying the same number of bits is contiguous, a property also experienced in practice and referred to as 'grouping' in ITU G.9991 \cite{itu2019high}. 
Our acceleration is inspired by the observation that  within such a group of subcarriers, the subcarrier at lowest frequency will need the least amount of extra power to carry one more bit.
Therefore, we only have to check a very limited number of candidate subcarriers, namely the first subcarrier that carries $b$ bits, for adding an extra bit during all iterations during bit loading. 
This approach significantly accelerates the HH algorithm by reducing the search space per iteration from $K$ subcarriers to the maximum bit level $B$ loaded by subcarriers. In practical systems, $B$ is small and unlikely to exceed 12 (4K-QAM).


We start with reserving memory $\tilde{\sigma}_x^2(k)$, $\Delta \tilde{\sigma}_x^2(k)$ and $b(k)$ to track the accumulated power, the incremental power required for an additional bit, and the bit level on each subcarrier $k$, respectively.
For a low-pass GNR, the algorithm assigns the first bit to the subcarrier at the lowest frequency, and updates $\tilde{\sigma}_x^2(k)$, $\Delta \tilde{\sigma}_x^2(k)$ and $b(k)$ for $k=1$. 

To short-list candidate subcarrier for bit loading, 
we create a lookup table $L(b)$ that indexes the lowest-frequency subcarrier that carries $b$ bits.
If the bit level $b$ is not used, we set the null table index for this bit level as $L(b)=0$ .
{In every iteration, when the bit loading of the selected subcarrier $k^*$ increases by one bit to $b(k^*)$, only two bit levels in the lookup table are affected, namely $L(b(k^*))$ and $L(b(k^*) - 1)$.}
For bit level $b(k^*)$, if this level is currently empty (thus $L(b(k^*))=0$), a new level is created by setting $L(b(k^*))$ to $k^*$. 
For instance, as shown in Fig.~\ref{fig:HH_table_1}, the algorithm updates the new bit level by setting $L(4)=1$ when $b(1)$ reaches 4. Otherwise, this bit level index remains unchanged, for example in Fig.~\ref{fig:HH_table_2}. 

Next, since $k^*$ is shifted from bit level $b(k^*)-1$ to $b(k^*)$, the tracking index for the previous bit level $L(b(k^*)-1)$ must be updated. 
For bit level $b(k^*)-1$, this index is shifted to the next subcarrier $k^*+1$ if it still belongs to the same bit level, where its power increment $\Delta \tilde{\sigma}_x^2({k^*+1})$ for the next bit is calculated. Accordingly, in Fig.~\ref{fig:HH_table_1}, we update $L(3)=2$ since $b(2)=3$. 
Otherwise, the index for this level is removed since no subcarrier is carrying $b(k^*)-1$ bits at this moment, 
as shown in Fig.~\ref{fig:HH_table_2} where we set $L(2)$ to null.

Within the updated lookup table $L$, select the new $k^*-$th subcarrier that needs the minimum power increment $\tilde{\sigma}_x^2{({k^*})}$ to load one bit, allocate this power to the subcarrier, and update this bit increment until the total power budget is exhausted.

\begin{algorithm}[ht]
\caption{Accelerated HH algorithm ($\text{HH}_\text{A}$)}
\label{alg:HH}
\begin{algorithmic}[1]
\STATE \textbf{Initialize:} $\mathrm{GNR}(f)$, maximum modulation frequency supported by the chipset $f_{\text{chip}}$, subcarrier number $K$, signal variance constraint $\sigma_x^2$, total allocated power $\tilde{\sigma}_{x,\text{tot}}^2 \leftarrow 0$, power allocated to all subcarriers $\tilde{\sigma}_x^2 \leftarrow 0_{1\times K}$, bit level of all subcarrier $b\leftarrow 0_{1\times K}$
\STATE Set $\Delta_{\mathrm{B}} = f_{\text{chip}}/K$, $W(\cdot) = \Delta_{\mathrm{B}}\Gamma \mathrm{GNR}^{-1}(\cdot)$;
\STATE $k^* \leftarrow 1$;
\STATE Update $\tilde{\sigma}_x^2({k^*})$, $\tilde{\sigma}_{x,\text{tot}}^2$, $b(k^*)$, and $\Delta \tilde{\sigma}_x^2({k^*})$;

\WHILE{true}
\IF{$L(b(k^*)) = 0$} 
    \STATE $L(b(k^*)) \leftarrow k^*$; \hfill (new bit level)
\ENDIF

\IF{$k^* < K$ \textbf{and} $b(k^*+1) = b(k^*)-1$}
    \STATE $L(b(k^*)-1) \leftarrow k^* + 1$; \hfill (update old bit level)
    \STATE $\Delta \tilde{\sigma}_x^2({k^*+1}) \leftarrow \tilde{\sigma}_x^2({k^*+1}) + W(f_{k^*+1})$;
\ELSE
    \STATE $L(b(k^*)-1) \leftarrow 0$;  \hfill (remove old bit level)
\ENDIF
\STATE Search $k^* \leftarrow \arg\min\limits_{k \in L(L \neq 0)} \Delta \tilde{\sigma}_x^2(k)$;

\IF{$\tilde{\sigma}_{x,\text{tot}}^2 + \Delta \tilde{\sigma}_x^2({k^*}) \ge \sigma_x^2$}
    \STATE \textbf{break};
\ENDIF

\STATE Update $\tilde{\sigma}_x^2({k^*})$, $\tilde{\sigma}_{x,\text{tot}}^2$, $b(k^*)$, and $\Delta \tilde{\sigma}_x^2({k^*})$;

\ENDWHILE
\end{algorithmic}
\end{algorithm}


\subsection{Complexity}
We characterize the computational complexity of the algorithm using both the big-O function $\mathcal{O}(\cdot)$ and the floating-point operations (FLOPs). 
In Algorithm \ref{alg:newton}, each Newton iteration for updating $f_\mathrm{max}$ requires $K$ operations to compute the total power. Each operation involves a $\mathrm{GNR}(f)$ evaluation with a complexity of $O(M+N)$, where $M$ and $N$ are the number of zeros and poles, respectively. In case the algorithm converges with $i$ iterations, the asymptotic complexity is $\mathcal{O}(K(M+N))$ since the convergence is usually fast with a small iteration number.
We quantify the computational workload as $(6i K+6K-5)(M+N) + i(13-8K)+2+K$ FLOPs. This complexity is predominantly determined by subcarrier number and the GNR model order, i.e. the number of poles and zeros.
For the $\text{HH}_\text{A}$, a (near-) optimal performance can be achieved with $R/\Delta_{\mathrm{B}}$ iterations. 
The  computational load per iteration is reduced by $(K-L)$ FLOPs through the use of a lookup table, which shrinks the set of candidate subcarriers.
In Section~\ref{sec:simu}, we compare the FLOPs of both algorithms with respect to the number of subcarriers, the GNR model order, and the throughput.

\section{Demonstration Using Experimental and Simulation Data}
\label{sec:simu}
In this section, we validate the pole-zero GNR model against the frequency responses of transceiver components from experimental measurements and simulations. The analytical throughput from \eqref{eqn:arctan_mzero_npole} is compared with that estimated from experimental data. Furthermore, the two algorithms developed in Section~\ref{sec:algorithm} are evaluated based on the resulting optimized signal spectrum, and the algorithm computational complexity.

\begin{table}[tb]\hspace{5pt}
\renewcommand{\arraystretch}{1.3}
\caption{Parameters from Experiments and Simulations}
\centering
\begin{tabular}{lclc}
\hline
Parameter &	  Symbol &	Value &	  Unit \\
\hline
LED 1st Pole & $f_{t,p1}$ & $2.3$ & \si{\mega\hertz} \\
LED 2nd Pole & $f_{t,p2}$ & $9.4$ & \si{\mega\hertz} \\
Phosphor Pole & $f_{t,p3}$ & $3.1$ & \si{\mega\hertz} \\
LED Zero & $f_{t,z}$ & $14.5$ & \si{\mega\hertz} \\
LED efficiency	 &$\eta_T$ & 	0.9 &  \si{\watt\per\ampere} \\
Propagation Gain & $h_0$ & $1\times10^{-5}$ & - \\
Receiver Pole & $f_{r,p}$ & $100$ & \si{\mega\hertz} \\
Receiver Zero & $f_{r,z}$ & $430$ & \si{\mega\hertz} \\
Noise Uplift Frequency & $f_{N}$ & $3.5$ & \si{\mega\hertz} \\
PD capacitance & $C_S$ & 130 & \si{\pico\farad} \\
PD responsivity	 &$\eta_R$ & 	0.5 &  \si{\ampere\per\watt}\\
Feedback Resistor	 &$R_F$ & 	100 &  \si{\ohm}\\
Modulation gap	 &$\Gamma$ & 	$6.06$ & \si{\decibel} \\
Noise Spectral Density & $N_0$ & $4.4\times10^{-18}$ & \si{\square \volt \per \hertz} \\
\hline
\end{tabular}
\label{tb:1}
\end{table}
For the transmitter, we consider the experimental setup in \cite{led_ho6} where a yellow phosphor-coated blue LED is driven by a bias-T modulator. The frequency response of an LED CXB1830-0000-000N0BV265E from Cree is measured, and a transfer function is derived with parameters (zeros and poles frequencies) estimated from the measurement data. The normalized frequency response of the LED and phosphor is characterized by a three-pole, one-zero function as
\begin{equation}
    |H_\mathrm{TX}(f)|^2 = \eta_T^2\frac{(1+f^2/f_{t,z}^2)}{\prod_{n=1}^3(1+f^2/f_{t,pn}^2)}.
    \label{eqn:tx model}
\end{equation}
We assume a LoS optical channel with a constant propagation gain $H_\mathrm{prop}(f)=h_0$. For the receiver, we consider a front-end with a large-area silicon PIN PD (S3204-08) connected to a TIA built with an off-the-shelf op-amp (AD8099) and a \SI{100}{\ohm} feedback resistor $R_F$. The front-end circuit is simulated on LTspice, which generates the gain response and the output noise spectral density of the circuit. The receiver response is modeled according to the op-amp datasheet and circuit parameters, resulting in a 5-pole, 4-zero transfer function as
\begin{equation}
    |H_\mathrm{RX}(f)|^2 = (\eta_R R_F)^2\frac{(1+f^2/f_{r,z}^2)^4}{(1+f^2/f_{r,p}^2)^5}.
    \label{eqn:rx model}
\end{equation}
The noise spectral density $S_N(f)$ has a noise power level at low frequency (flicker noise ignored) of $N_0$. The uplift frequency corner $f_N$ is calculated based on \cite{receiver_noise}, which shows a good match to the simulation. At high frequency, the noise is attenuated by the amplifier with the same frequency roll-off as \eqref{eqn:rx model}. We express the noise spectral density $S_N(f)$ as
\begin{equation}
    S_N(f)= N_0\frac{(1+f^2/f_{N}^2)(1+f^2/f_{r,z}^2)^4}{(1+f^2/f_{r,p}^2)^5}.
    \label{eqn:noise model}
\end{equation}
When calculating the GNR, the 5-pole, 4-zero frequency response caused by the amplifier cancels out in both the numerator \eqref{eqn:rx model} and the denominator \eqref{eqn:noise model}. Therefore, the $\mathrm{GNR}(f)$ of this OWC link can be expressed as
\begin{equation}
\begin{split}
    \mathrm{GNR}(f) & \hspace{-3pt}= \hspace{-3pt} \frac{|H_\mathrm{TX}(f)|^2 |H_\mathrm{prop}(f)|^2 |H_\mathrm{RX}(f)|^2}{S_N(f)}\\
    &\hspace{-40pt}=\hspace{-3pt}\eta_T^2 h_0^2 \eta_R^2 R_F^2 \frac{(1+f^2/f_{t,z}^2)}{N_0(1+f^2/f_{N}^2)\prod_{n=1}^3(1+f^2/f_{t,pn}^2)}.
\end{split}
\label{eqn:gnr_all}
\end{equation}
The corner frequency of the zeros and poles, as well as the parameters that determine $\mathrm{GNR}_0$ are illustrated in Table \ref{tb:1}. 

\begin{figure*}[tb]
    \centering
    \captionsetup[subfloat]{labelfont=scriptsize,textfont=scriptsize}
    \subfloat[Gain (\si{\ohm}) \label{fig:gain}]{%
        \includegraphics[width=.33\linewidth]{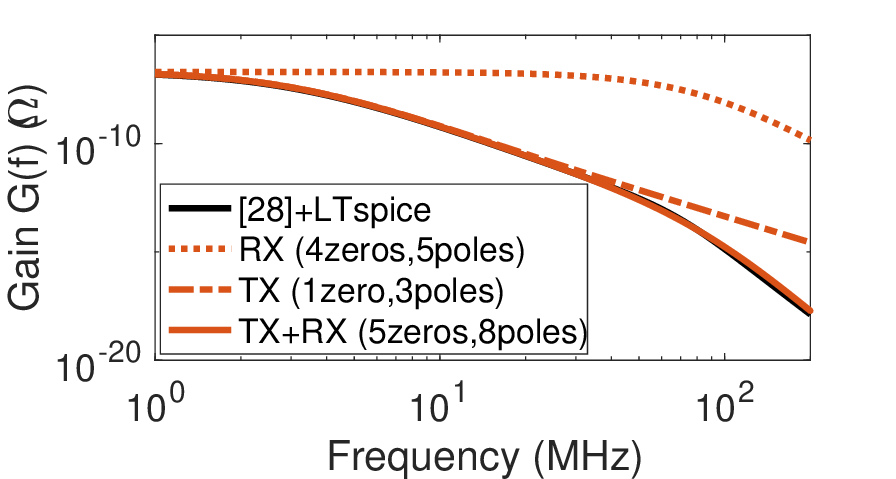}
    } 
    \subfloat[Output Noise PSD (\si{\square\volt\per\hertz})\label{fig:noise}]{%
        \includegraphics[width=.33\linewidth]{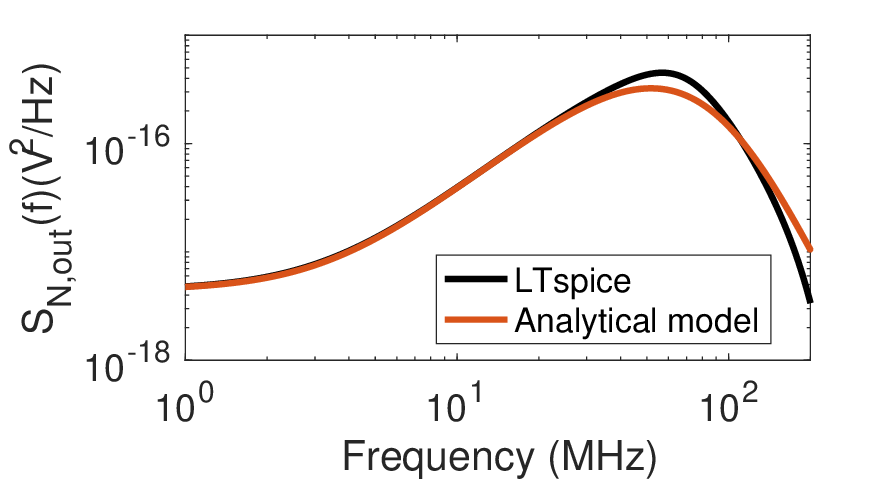}    } 
    \subfloat[GNR$(f)$ (\si{\ohm\per\volt\squared}/\si{\hertz}) \label{fig:GNR}]{%
        \includegraphics[width=.33\linewidth]{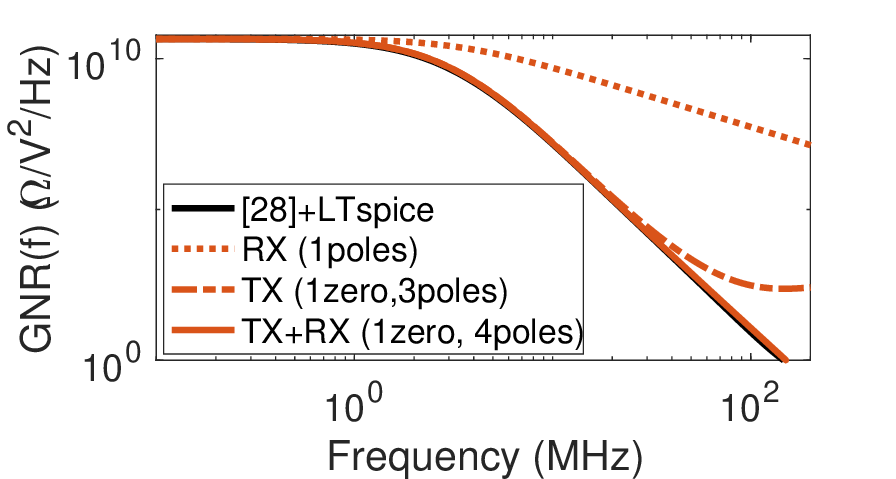}
    } 
    \caption{Characteristics of signal response, noise PSD, and GNR: A comparison of the fitted pole-zero functions (red curves) against the reference data from \cite{led_ho6} and LTspice simulations (black curves).}
    \label{fig:simu}
    \vspace{-5pt}
\end{figure*}
Fig.~\ref{fig:simu} compares the analytical pole-zero model with the experimental and simulation results. 
Fig.~\ref{fig:gain} shows the gain response from the simulation, and the frequency response model of the whole OWC link, i.e., including the transmitter \eqref{eqn:tx model} and the receiver \eqref{eqn:rx model} responses.
The DC gain of the responses are unified as $\eta_T^2 h_0^2 \eta_R^2 R_F^2$, so that only the differences in their frequency characteristics are pronounced.
The results show that the transmitter introduces the main bandwidth limitation since the LED has a relatively low speed. The total frequency response is dominated by the transmitter until around \SI{30}{\mega\hertz}, where the frequency response of PD and TIA starts to roll off. At this time, the total frequency response deviates from the transmitter function with a higher order decay.
The noise PSD is properly modeled with uplift frequency $f_N$ as is shown in Fig.~\ref{fig:noise}, while at higher frequencies a slight mismatch shows due to a complicated op-amp frequency behavior. The overall GNR in Fig.~\ref{fig:GNR} shows that an analytical model that incorporates all stages of the OWC link better matches the simulation, while considering only the TX or RX introduces significant errors.


\begin{figure}[tb]
    \centering
    \captionsetup[subfloat]{labelfont=scriptsize,textfont=scriptsize}
    \subfloat[\label{fig:throughput_f}]{%
        \includegraphics[width=.75\linewidth]{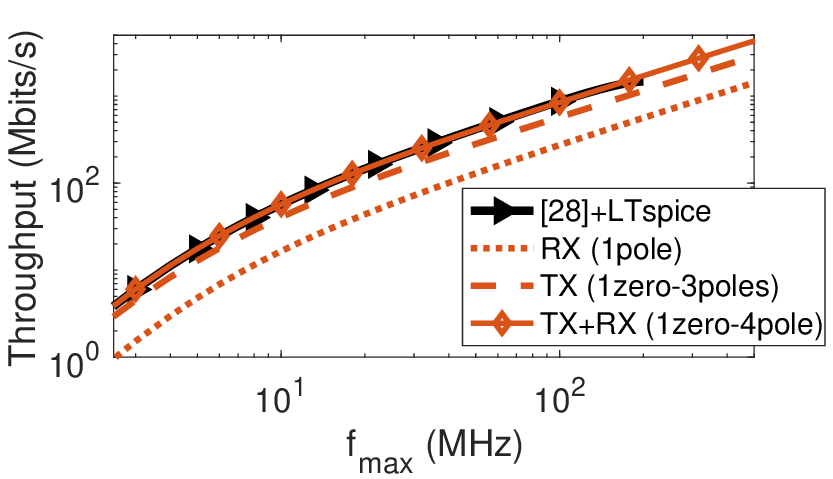}
    }\\
    \subfloat[\label{fig:throughput_sigma}]{%
        \includegraphics[width=.75\linewidth]{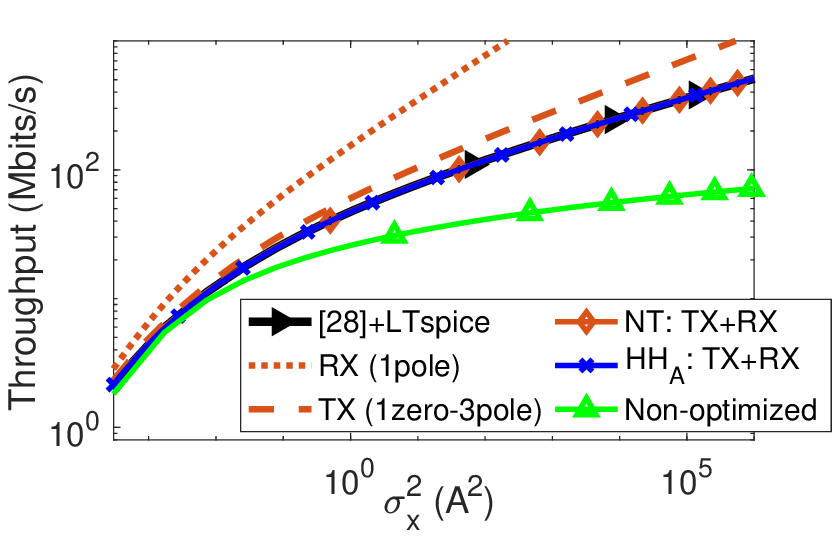}
    }
    \caption{Throughput estimation for the proposed pole-zero analytical GNR model and LTspice simulations (black curves) as a function of: (a) maximum modulation frequency $f_{\max}$, and (b) signal variance $\sigma_x^2$. In (b), throughput optimization is performed using NT and $\text{HH}_\text{A}$. The green curve in (b) represents the throughput of an unoptimized (flat) signal spectrum with $f_{\max} = f_{t,p1}$.}
    \label{fig:R}
     \vspace{-5pt}
\end{figure}
We show the estimated throughput optimized via waterfilling, with respect to the maximum modulation frequency in Fig.~\ref{fig:throughput_f}, and to the signal variance in Fig.~\ref{fig:throughput_sigma}, respectively.
For the throughput calculated from the GNR that includes all stages, the result matches well with the throughput estimated from the simulation. However, models that account for only the TX or RX components yield an inaccurate throughput estimation by neglecting the cumulative bandwidth constraints. 
As illustrated in Fig.~\ref{fig:throughput_f}, neglecting higher-order roll-off leads to an insufficient throughput estimation for a given $f_\mathrm{max}$. 
For a limited $\sigma_x^2$, the throughput is overestimated when bandwidth limitations are not fully considered, as shown in Fig.~\ref{fig:throughput_sigma}. 
We further compare the throughput optimized by NT (Algorithm~\ref{alg:newton}) and the $\text{HH}_\text{A}$  (Algorithm~\ref{alg:HH}), with the baseline scheme where signal power is uniformly distributed and modulated within the flat region of $\mathrm{GNR}(f)$, as shown by the green curve in Fig.~\ref{fig:throughput_sigma}. 
The performance is close to \eqref{eqn:arctan_mzero_npole} for weak signals. As the power budget increases, the waterfilling–optimized throughput shows a significant advantage over the green curve, since a larger modulation bandwidth is used and the signal spectrum is adapted to the channel frequency characteristics.

\begin{figure}[tb]
    \centering
    \captionsetup[subfloat]{labelfont=scriptsize,textfont=scriptsize}
     \subfloat[ \label{fig:algorithm_PSD}]{%
        \includegraphics[width=.9\linewidth]{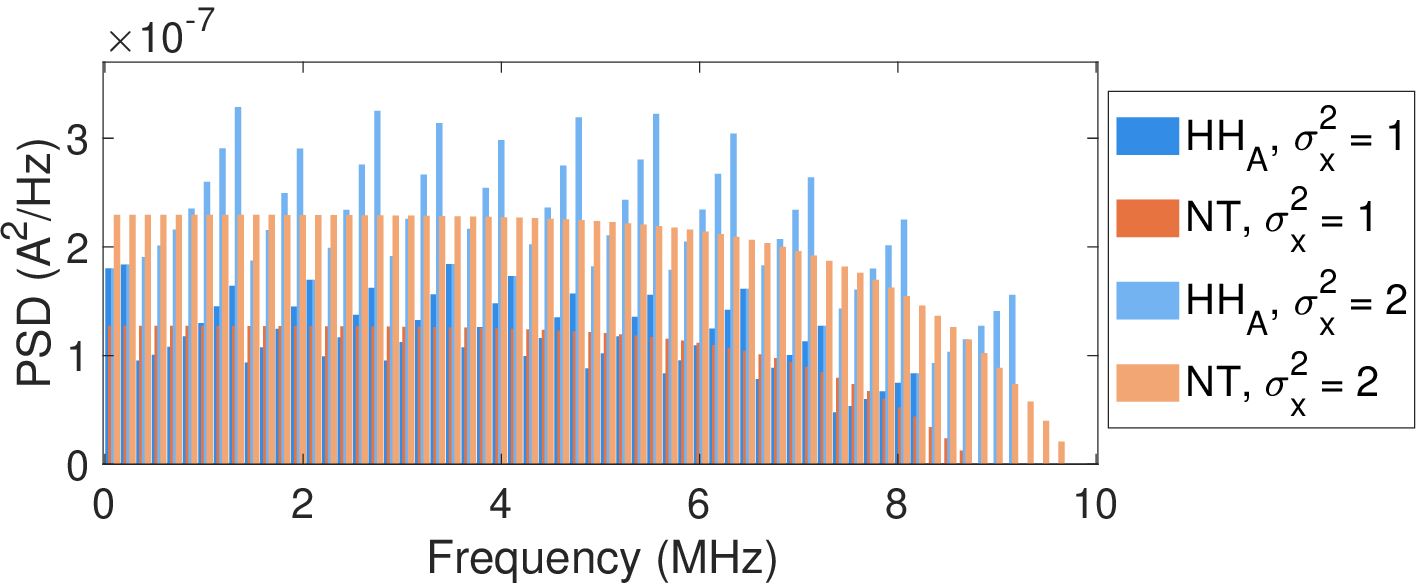}
    } \\
    \subfloat[ \label{fig:algorithm_complexity}]{%
        \includegraphics[width=\linewidth]{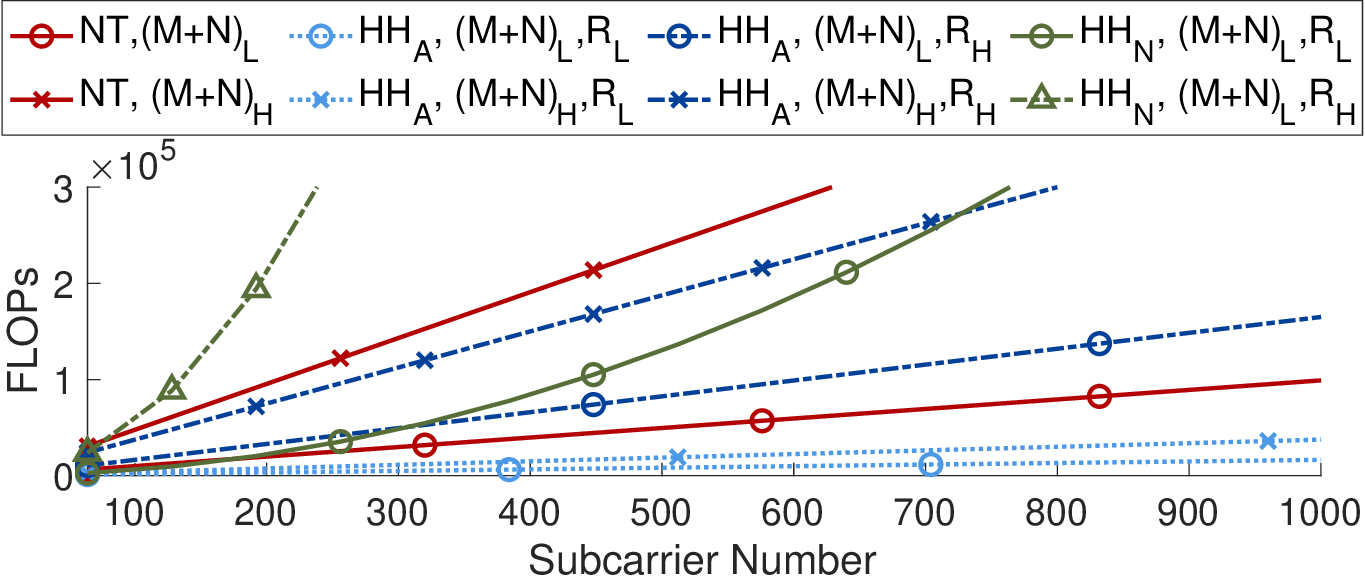}    } 
    \caption{(a) Result of optimized signal PSD optimized by NT and $\text{HH}_\text{A}$, for varies $\sigma_x^2$. Subcarriers number $K=64$, $f_\text{chip}=$\SI{200}{\mega\hertz}, and (b) FLOTs of NT, $\text{HH}_\text{A}$ and non-accelerated HH in \cite{10295463} ($\text{HH}_\text{N}$) vs. $K$, for low and high GNR order $(M \hspace{-2pt} + \hspace{-2pt} N)_L \hspace{-2pt} = \hspace{-2pt}3$, $(M \hspace{-2pt} + \hspace{-2pt} N)_H \hspace{-2pt} = \hspace{-2pt} 10$ and different rate $R_L=$\SI{100}{\mega\bit\per\second} and $R_H=$\SI{1}{\giga\bit\per\second}.}
    \label{fig:algorithm}
    \vspace{-5pt}
\end{figure}
Fig.~\ref{fig:algorithm_PSD} shows the optimized signal spectrum from NT and $\text{HH}_\text{A}$, for the channel modeled in \eqref{eqn:gnr_all}. 
While NT is formulated for continuous rate optimization as in \eqref{eqn:arctan_mzero_npole}, a portion of the power allocated to each subcarrier may remain effectively 'unused' because it is insufficient to support an additional integer bit, as illustrated by the mismatch of the signal spectrum optimized from the two algorithms.
The computational cost, quantified in FLOPs, is compared in Fig.~\ref{fig:algorithm_complexity}. 
As the GNR model order increases, the computational complexity of NT rises significantly, as its per-iteration operations are highly sensitive to the number of poles and zeros. In contrast, the $\text{HH}_\text{A}$ exhibits high efficiency in low data rate scenarios. However, its performance degrades as the data rate increases which requires more bit-loading iterations. 
In other words, NT is faster for spectrally lower-order GNRs (simpler expressions), while the $\text{HH}_\text{A}$ is faster and more efficient for low power budgets because fewer bits per OFDM block require fewer iterations.
Furthermore, we show the complexity reduction by applying the lookup table, in comparison to previous work on HH bit loading for bandwidth-limited OWC \cite{10295463}, represented by the green curves $\text{HH}_\text{N}$ in Fig. \ref{fig:algorithm_complexity}. The results indicate substantial FLOPs saving with the acceleration, which is further pronounced with the increasing number of subcarriers.


\section{Conclusions}
\label{sec:conclusion}
This paper presented optimization solutions for enhancing the throughput of DCO-OFDM OWC systems subject to the cascade of bandwidth-limited components. 
We emphasize the accuracy of the proposed spectral GNR model for the complete system by explicitly assessing the imprecision on the estimation that occurs when the bandwidth limitations are only partially considered, as is commonly done in the existing literature.
With the proposed algorithms, including a significant acceleration on the existing HH algorithm, the optimal power spectrum can be efficiently allocated for a given power budget. Our results have shown significant throughput enhancement compared to the modulation scheme where the signal is modulated only within the flat region of the spectral GNR. Moreover, the closed-form expression of achievable throughput is particularly practical for OWC systems, as the time-invariant nature of transceiver components ensures stable and accurately estimable CSI.









\begingroup
\fontsize{8pt}{9.6pt}\selectfont
\bibliographystyle{./bibliography/IEEEtran}
\bibliography{IEEEabrv,./bibliography/reference}
\endgroup

\end{document}